\documentclass[aip, jcp, jmp, amsmath,amssymb, preprint]{revtex4-1}
\usepackage{amsmath}
\usepackage{latexsym}
\usepackage{graphicx}
\usepackage{color}
\usepackage{fancyhdr}
\usepackage{hyperref}

\begin{document}
\title{An Ultra Low Noise Telecom Wavelength Free Running Single Photon Detector Using Negative Feedback Avalanche Diode}
\author{Zhizhong Yan}
\email{zyan@iqc.ca} \affiliation{Institute for Quantum Computing,
University of Waterloo, 200 University Avenue W, Waterloo N2L 3G1,
Canada}

\author{Deny R. Hamel}
\affiliation{Institute for Quantum Computing, University of
Waterloo, 200 University Avenue W, Waterloo N2L 3G1, Canada}

\author{Aimee K. Heinrichs}
\affiliation{Institute for Quantum Computing, University of
Waterloo, 200 University Avenue W, Waterloo N2L 3G1, Canada}

\author{Xudong~Jiang}
\affiliation{Princeton Lightwave Inc., 2555 US Route 130 S.,
Cranbury, NJ 08540  USA}

\author{Mark~A.~Itzler}
\affiliation{Princeton Lightwave Inc., 2555 US Route 130 S.,
Cranbury, NJ 08540  USA}

\author{Thomas~Jennewein}\email{thomas.jennewein@uwaterloo.ca}
\affiliation{Institute for Quantum Computing, University of
Waterloo, 200 University Avenue W, Waterloo N2L 3G1, Canada}

\begin{abstract}
It is challenging to implement genuine free running single photon
detectors for the 1550 nm wavelength range with simultaneously high
detection efficiency (DE), low dark noise, and good time resolution.
We report a novel read out system for the signals from a negative
feedback avalanche diode
(NFAD)\cite{Jiang_SPIE_09_NFAD,Jiang_SPIE_11_NFAD,Itzler_SPIE_10}
which allows useful operation of these devices at a temperature of
193 K and results in very low dark counts ($\sim$ 100 CPS), good
time jitter ($\sim$ 30 ps), and good DE ($\sim$10 $\%$). We
characterized two NFADs with a  time correlation method using
photons generated from  weak coherent pulses (WCP) and photon pairs
produced by spontaneous parametric down conversion (SPDC). The
inferred detector efficiencies for both types of photon sources
agree with each other. The best noise equivalent power of the device
is estimated to be $8.1 \times 10^{-18}~\mathrm{W \cdot Hz}^{-1/2}$,
more than 10 times better than typical InP/InGaAs SPADs show in free
running mode. The afterpulsing probability was found to be less than
$0.1\%$ per ns at the optimized operating point. In addition, we
studied the performance of an entanglement-based quantum key
distribution (QKD) using these detectors and develop a model for the
quantum bit error rate (QBER) that incorporates the afterpulsing
coefficients. We verified experimentally that using these NFADs it
is feasible to implement QKD over 400 km of telecom fibre. Our NFAD
photon detector system is very simple, and is well suited for
single-photon applications where ultra-low noise and free-running
operation is required, and some afterpulsing can be tolerated.
\end{abstract}

\maketitle
\section{Introduction}
Single photon detectors (SPD) are important devices in many research
fields, such as optical quantum computing \cite{KLM_Nature_01},
quantum cryptography \cite{BB84, Gisin_RevModPhy_02_QKD},
spectroscopy, fluorescence lifetime measurements
\cite{Stellari_TED_01}, and light detection and ranging (LIDAR)
\cite{Williams_SPIE_06_linearMode}. The telecom wavelength range
around 1500 nm is of particular interest for various reasons. For
example, quantum key distribution (QKD) via optical fiber must
operate at the lowest photon transmission loss at 1550 nm in order
to achieve the longest distance.

SPDs can be operated in two distinct modes: gated
\cite{Tosi_SPIE_06} or free running \cite{Thew_APL_07,
Warburton_09_APL, Zhang_SPIE_10, Zhao_APL_07, Cheng_OptExpress_11}.
Free running detectors are very important in certain applications,
particularly where photons arrive in a random fashion, rather than
synchronized on a regular clock time basis. For example, this is the
case for photon pairs produced by spontaneous parametric down
conversion (SPDC) with a continuous wave pump.

In the telecom wavelength range there exist several good candidates
for free running SPDs, including photomultiplier tubes (PMT)
\cite{Horsfield_RSI_10}, passively quenched InP/InGaAs SPADs
\cite{Itzler_SPIE_07}, shallow junction silicon SPADs using
frequency upconversion \cite{Albota_04_upconversion}, transition
edge sensors (TES)\cite{TES:Lita:08}, and superconducting nanowire
single photon detectors (SNSPDs) \cite{Goltsman_APL_01}.

The SNSPDs currently have the best overall performance, such as low
darkcount rates (DCR $<$ 100 Hz) and good timing jitter ($<$ 65 ps).
However, owing to their superconducting temperature requirements
(under 4.2 K), cryogenic systems are required. Moreover, the
detection efficiency (DE) of SNSPDs is dependent on the polarization
of the input photons\cite{zyan_TAS_07, Anant_OptExpress_08}. On the
other hand, negative feedback avalanche diodes (NFADs), a recently
developed type of InP/InGaAs APD (Princeton Lightwave Inc), can be
operated in free running mode without cryogenic temperatures. These
devices possess an integrated passive quenching resistor that
minimize the amount of avalanche charges produced by a photon
detection (another approach to self-quenching in InP-based single
photon APDs, which employs epitaxial barriers as a negative feedback
mechanism \cite{Zhao_APL_0701, Zhao_APL_08}). Furthermore, these APD
detectors have no DE dependency on the input photon polarization.

The darkcount rate and timing jitter of NFAD detectors arise from
some intrinsic material properties and the device structure. Their
darkcount rates are induced by the size of their avalanche region,
as well as the trap-assisted tunneling
effect\cite{Jiang_SPIE_11_NFAD} (TAT). Timing jitter is generally
dominated by the stochastic nature of the avalanche breakdown
process. However, the external operation setup can be optimized to
achieve a high signal-to-noise ratio (SNR) and good time resolution.
We developed a novel readout system for the NFADs, which features
(a) very low operating temperature; and (b) minimal avalanche
charges required to trigger the readout circuit. This allows our
system to achieve ultra low darkcount rates.

The significant features of our readout method are: first, the NFAD
operates in a genuine free running mode; there is no active
quenching circuitry involved to impose additional deadtime (which is
usually needed to suppress the afterpulsing effect and to lower
darkcount rates); Second, a pulse transformer is employed to isolate
the NFAD bias circuitry from the high speed RF amplifiers to read
out the photon electric signals. Instead of conventional capacitive
coupling, we use inductive coupling, so that only the variation of
the avalanche current is coupled to the external amplifier. This
dramatically improves the sensitivity of the readout.

In this work two different NFAD devices are
studied\cite{Itzler_SPIE_10}. Detector I/NFAD 1 (model no. E2G6) has
a feedback resistor of 1.1 M$\Omega$ and a diameter of 22 $\mu$m.
Detector II/NFAD 2 (model no. E3G3) has the same nominal feedback
resistor, but a larger device diameter of 32 $\mu$m.

The paper is organized as follows: we start with the experimental
setup and device details. Then the ``Simulation Program with
Integrated Circuit Emphasis" (SPICE) model of our readout circuit
including the NFAD equivalent circuit is simulated and compared with
the experimental results. This is followed by a measurements of the
DE, which is later verified using two NFADs to detect photon pairs
generated by SPDC. Next the results from the afterpulsing, deadtime,
and timing jitter measurements are presented. Lastly, we study the
feasibility of using our NFADs to perform entangled photon QKD over
distances of more than 400 km.
\section{The Detector System}
\subsection{Setup}
\begin{figure}
\includegraphics[width= 0.8 \textwidth]{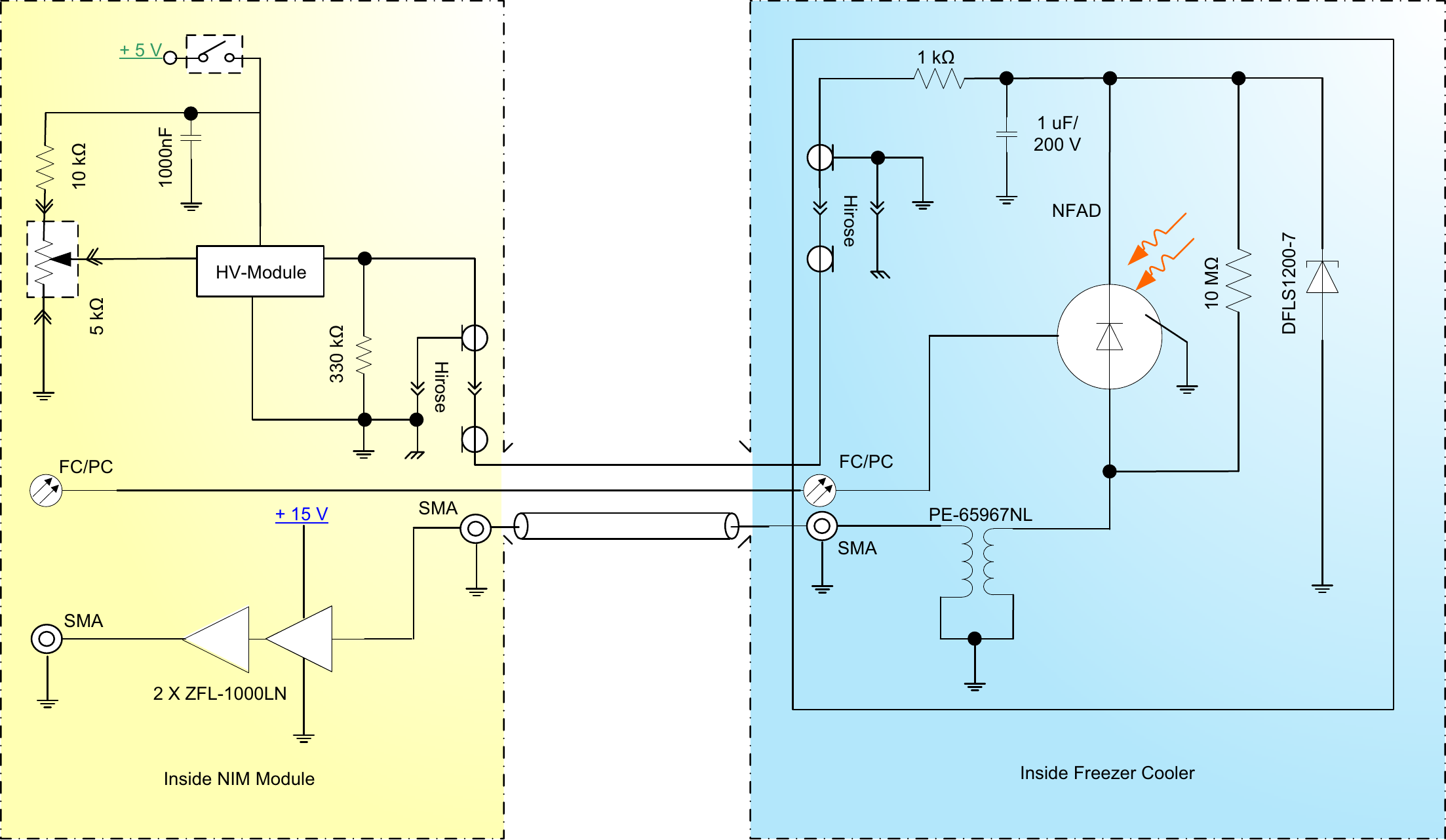}
\caption{The system design diagram of the entire NFAD readout
circuits and connections.  All the components in the right panel are
inside a deep freezer.  The ones in the left panel are at room
temperature. SMA is the high frequency connector; FC/PC is the
single mode fiber connector; Hirose is the NFAD high voltage cable
connector.}\label{fig:System_Design}
\end{figure}
Fig. \ref{fig:System_Design} depicts the detailed design of our
in-house developed readout circuit. The entire system is comprised
of two parts. The NFAD and its bias circuitry is inside a deep
freezer (Global Cooling, ULT-25) with a lowest base temperature of
187~K. At this base temperature, the detector operating temperature
is 193 K. The primary coil of the pulse transformer (Pulse
Electronics) is connected between the system ground and the cathode
of the NFAD. The secondary coil of the transformer couples to a 50
$\Omega$ transmission line (TL). A bypass resistor of 10 M$\Omega$
and a Schottky diode serve to protect the NFAD from unexpected
voltage surges or spikes.

The rest of the readout electronics, including high voltage modules
and the wide band RF amplifier, are in a nuclear instrumentation
module (NIM) at room temperature. The high voltage control module
(Matsusada) supplies variable bias voltage to the NFAD, and its
output value is controlled by a multi-turn linear potentiometer. The
bias voltage spans from 0 to 80 V. Two wide band RF amplifiers
(Mini-Circuits) have 40 dB gain in total to amplify the signals from
the pulse transformer via the TL.
\begin{figure}[h]
\includegraphics[width= 0.7 \textwidth]{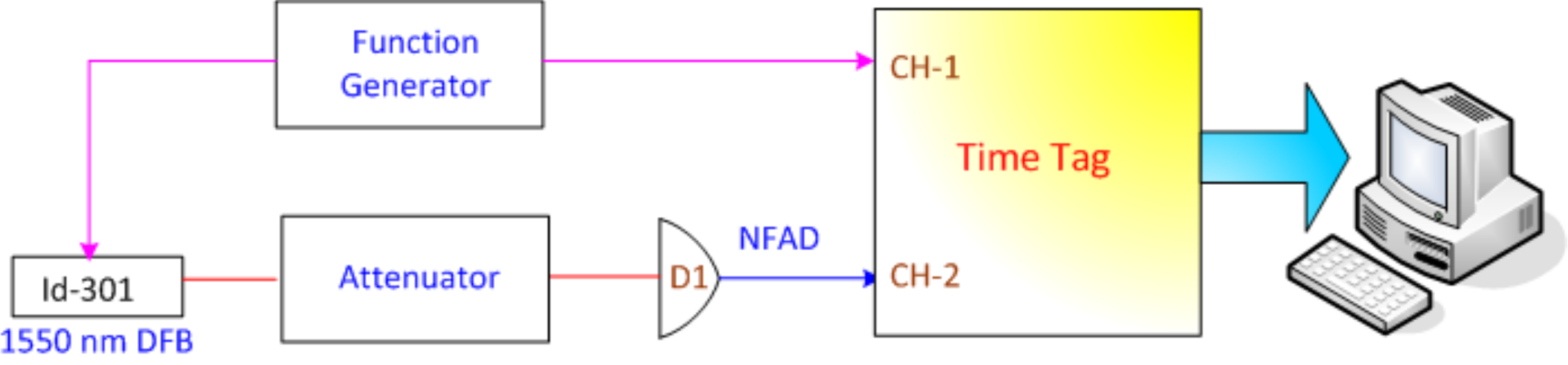}
\caption{Testing setup for characterization of the NFAD using the
time-correlated single photon counting method. Details in the
text.}\label{fig:Setup}
\end{figure}

The experimental setup to characterize the NFADs is based on
time-correlated single photon counting (TCSPC), see Fig.
\ref{fig:Setup}. Optical pulses are generated with an id-301 1550 nm
distributed feedback (DFB) laser, attenuated by an EXFO programable
attenuator (FVA-3100). The laser is triggered by a two-channel
function generator (Tektronix AFG3502). The attenuated weak coherent
pulses pass through the attenuator and are sent to the NFAD. The
time tag unit records detections and the synchronization signals,
and establishes their coincidences.
\subsection{The photon detection pulse responses}
\begin{figure}
\includegraphics[width=0.6 \textwidth ]{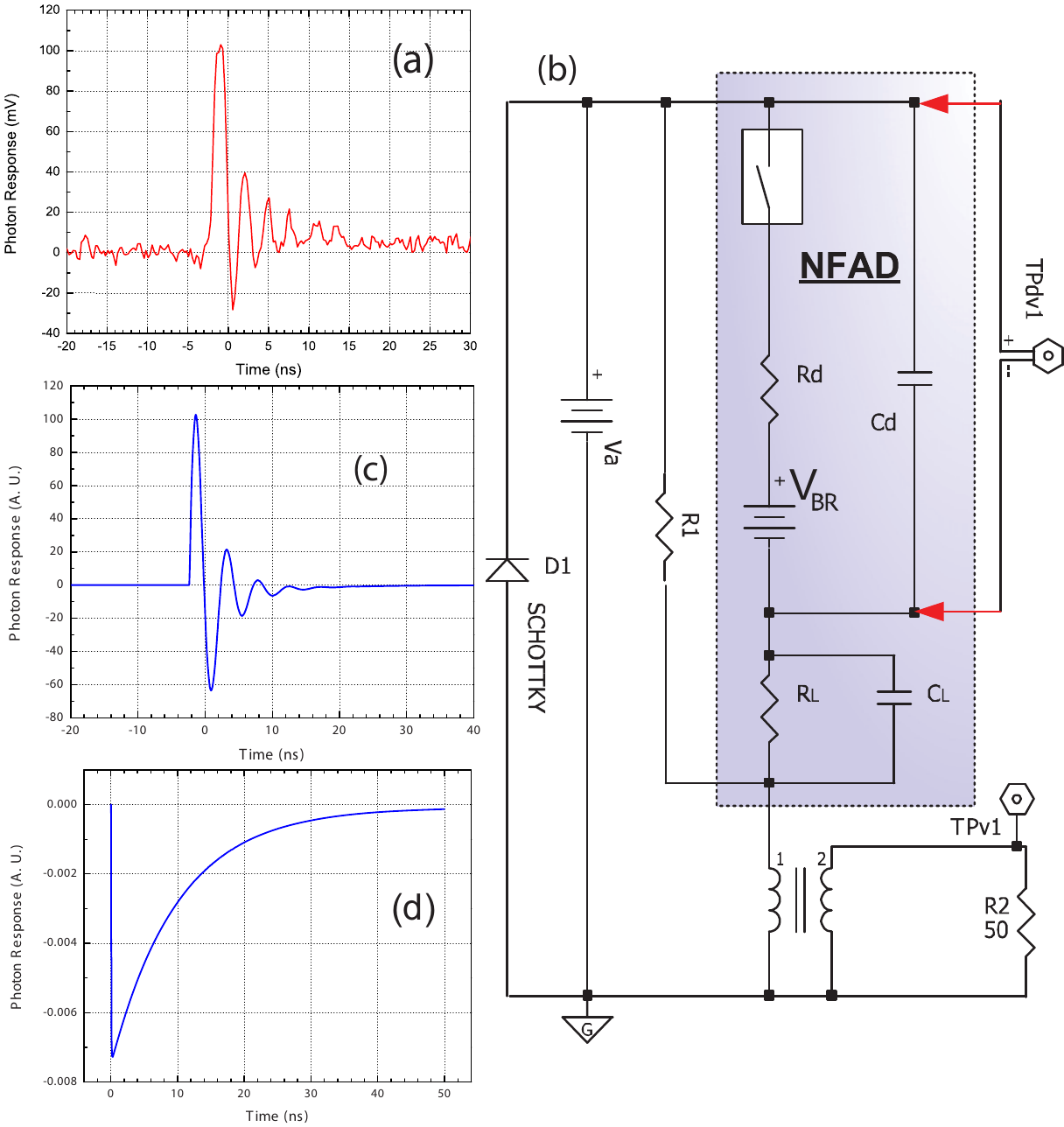}
\caption{(a) The experimental single photon response recorded with a
fast digital oscilloscope after the 40 dB gain amplification. (b)
The SPICE model for the NFAD read out circuit. The dash line square
highlights the NFAD equivalent circuit components; The testing
points TPv1 and TPdv1 represent photon detection response output and
intrinsic SPAD voltage drop, respectively; the voltage source VBR
represents the breakdown voltage of the NFAD. (c)-(d) are results of
the transient simulation of a single photon responses, before the
amplifier. (c) is the simulated response for the readout circuits
used in this work; (d) is the simulated response with the
conventional readout methods using a bias-tee in Ref.
\onlinecite{Jiang_SPIE_09_NFAD}. The model parameters for the bias
tee in the the conventional method are simplified from the ones
provided by its manufacturer.}\label{fig:SPICE_Model_Sim}
\end{figure}

The photon detection response is measured by a fast digital
oscilloscope. Fig. \ref{fig:SPICE_Model_Sim} (a) shows a typical
waveform  from an experimental measurement of the single photon
response. We also simulated the response using a SPICE model for the
equivalent circuit depicted in Fig. \ref{fig:SPICE_Model_Sim} (b)
and  the NFAD model based on the results of Hayat et al.
\cite{Hayat_NFAD_model_10}. The simulation allows us to estimate the
heating  induced by each photon detection event, as well as the
amount of avalanche charges coupled to the external readout circuit.
Both estimations can be achieved by simulating transient response
simulation of a single avalanche event.

The simulation results for the signal response of two readout
methods are illustrated in Fig. \ref{fig:SPICE_Model_Sim}: (c) shows
  the transient response based on our readout circuit design
(see Fig. \ref{fig:SPICE_Model_Sim} (b)); graph (d) is the result
for the conventional readout method shown in Ref.
\onlinecite{Jiang_SPIE_09_NFAD}. The conventional circuits transient
response has a persistent current plateau ranging from $\sim$10-100
ns and is proportional to the overbias voltage \cite{Itzler_SPIE09}.
In comparison, our main response voltage peak is less than 5 ns, and
is independent of overbias voltage. The slow variations of the
persistent current induce almost no voltage via the transformer to
the amplifier input. The simulation and experiment results agree
well for the leading period of about 5 ns long. The experimental
data exhibit a more complicated response, because the current SPICE
model only accounts for the main physical process.

In order to study the power consumption arising from the Joule
heating effect, we performed numerical integration over the product
of total power supply current and voltage. Two sources of heat
generation are identified: a constant background heat power of $\sim
5 \times 10^{ - 4}~\textrm{W}$ induced by bias and protection
circuitry (e.g. $10~\mathrm{M \Omega}$ resistor $\mathrm{R_1}$), and
an additional $4 \times 10^{ - 12}~\textrm{J}$ produced by every
detection.  Therefore, even at very high detection rates, the
heating caused by detections is negligible, compared to other
sources of heat, such as the background heat.

Moreover, it is of interest to analyze the net avalanche charge,
$Q_\mathrm{av}$, coupled to the external amplifier. We performed the
numerical integration over the simulated voltage across a
50~$\Omega$ load via $Q_\mathrm{av} = {1 \over {50}}\int_0^\tau
\mathrm{{V_{TPv1}}  \cdot dt}$. The computed result showed the
avalanche charge with inductive coupling is about $5 \times 10^{ -
16}$ C (corresponding to $3 \times 10^3$ electron charge~$e^-$).
Comparing this to the net avalanche charge for a conventional
readout circuit \cite{Jiang_SPIE_11_NFAD}, ours consumes at least
two orders of magnitude less avalanche charges to produce a
legitimate photon response pulse. This low coupling charge leads to
the improved sensitivity of our readout circuit, and contributes to
the low afterpulsing dark noise performance of the detector system.
\section{Detection Efficiency Characterization}
\subsection{Repetitive weak coherent states method}
We first use attenuated coherent states, prepared at regular
repetition rate, to characterize the detection efficiency of the
NFAD. We choose the period of the repetitive pulse ($T_0 = 100
\mu$s) to be much longer than the detector recovery time ($\tau =
50$ ns). The coherent state of each pulse is represented in the Fock
basis by $ \left| \alpha \right\rangle = e^{ - {{\left| \alpha
\right|^2 } \over 2}} \sum\limits_{n = 0}^\infty {{{\alpha ^n }
\over {\sqrt {n!} }}} \left| n \right\rangle $, where the average
photon number contained in each pulse is $ \mu  = \left| \alpha
\right|^2 $. Since
 detection events are identified by comparing the voltage level of an
 photon response pulse to a pre-setting threshold, we model
this type of detector as a so-called bucket detector
\cite{Jennewein_JMO_11}. The probability for click and no-click of
the bucket detector are described by $ \hat f_\mathrm{click}  =
\sum\limits_{n = 0}^\infty {1 - \left( {1 - \eta _D } \right)^n
\left| n \right\rangle } \left\langle n \right| $ and $ \hat
f_\mathrm{no\:click} = \sum\limits_{n = 0}^\infty  {\left( {1 - \eta
_D } \right)^n \left| n \right\rangle } \left\langle n \right| $
respectively.

In the experiment, we  measure the ratio of detected photon click
events, $R_{\det }$, within the 1~ns correlated time window, $\tau
_0$, versus the photon trigger repetition rate, $f_\mathrm{rep}$,
under a certain average photon number per pulse, $\mu$. We combine
the coherent quantum states to derive the detection efficiency,
$\eta_\mathrm{D} $ with the TCSPC method as:
\begin{equation}\label{eq:eta_D}
\eta _\mathrm{D}  = -{1 \over \mu }\ln \left( {{1 - {{R_\mathrm{det}
\over {f_\mathrm{rep} }}} \over {1 - d_\mathrm{B} \tau _0 }}}
\right),
\end{equation}
where the total non-photon detection events, $d_\mathrm{B}$, becomes
a summation of pure dark counts, $D$, and afterpulsing events
induced by photon detections, $d_\mathrm{afp}$, i.e. $d_\mathrm{B} =
D + d_\mathrm{afp}$. $\tau_0$ is the smallest time window in the
TCSPC counting system. Here we always use a 1~ns time window for all
measurements with the time tag unit.

Because the darkcount and repetition rates are both below 10~kHz,
and $\tau _0 = 10^{-9}$ s, we are able to use ${1 - d_\mathrm{B}
\tau _0 } \simeq 1 $ as a very good approximation in most of our DE
measurements. Thus, we have a practical relation to compute the DE:
\begin{equation}\label{eq:eta_approx}
\eta _\mathrm{D}  \simeq -{1 \over \mu }\ln \left( {1 -
{{R_\mathrm{det}} \over {f_\mathrm{rep} }}} \right).
\end{equation}
Here we assume the detector operator is known to behave as a bucket
detector. If one wants to characterize a completely unknown
detector, then a more general approach would be to perform the
detector tomography\cite{Lundeen_NatPhy_09, Akhlaghi_OptExp_11}.

\begin{figure}
\includegraphics[width= 0.8 \textwidth]{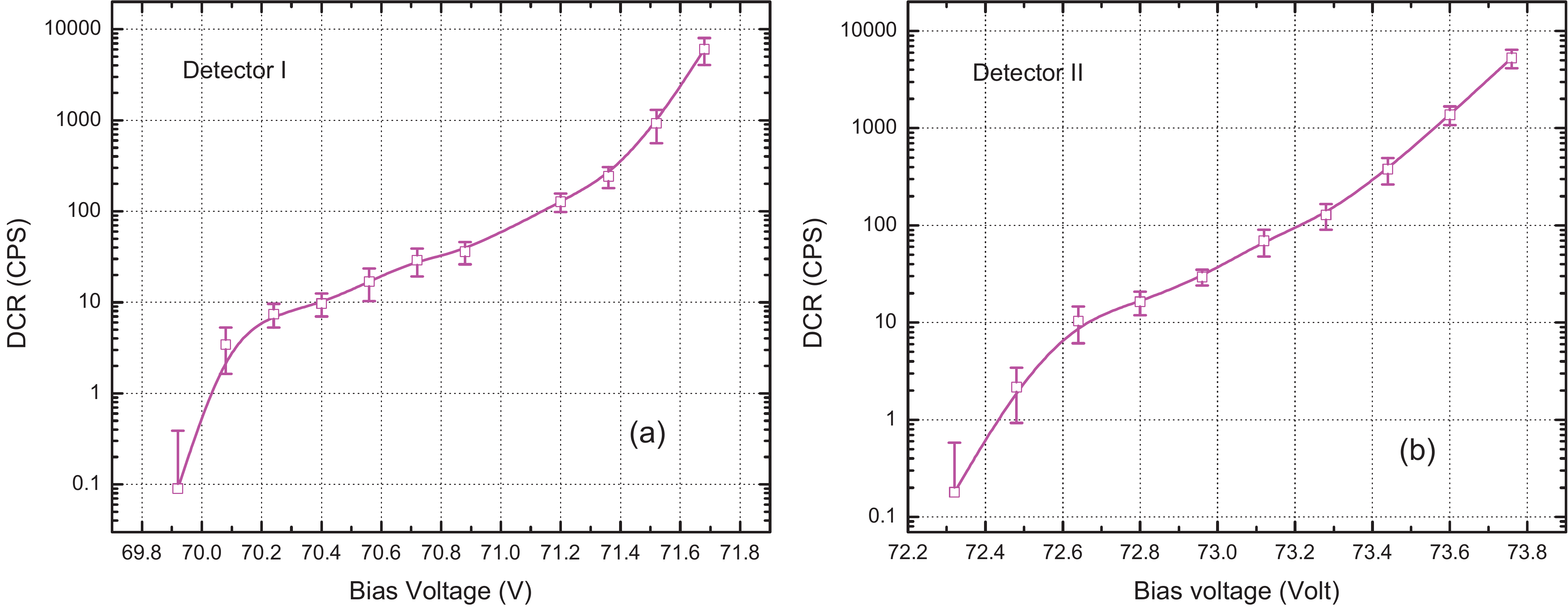}
\caption{Measurement of the darkcount rates for the two NFAD
devices. Both are measured at  193 K with the same discriminator
voltage settings.}\label{fig:Dark-Voltage}
\end{figure}

Prior to the detection efficiency (DE) measurements, we first
measure the darkcount rate for various bias voltages. Both detectors
exhibit a similar behavior of voltage to darkcount rate. Fig.
\ref{fig:Dark-Voltage} shows the trend of darkcount rate as a
function of input voltage. Over the range of overbias voltages
tested, the darkcount rate varies from 0.1 to 10,000 CPS.

\textcolor[rgb]{0,0.00,0.00}{The DE testing conditions are as
follows: the laser source, id-301, is triggered at 1 kHz with its
output attenuated such that the photon flux is determined to be 1
photon per pulse by a calibrated power meter \cite{Reply_1a}. We
then verify this attenuation using a commercial detector
\cite{Reply_1b}, id-201,
 set to 10~\% detection probability with a 20 ns gate
and a 40 ns deadtime.  We verify that the ratio of detected photon
counts vs. total triggering count reaches $9.5 \%$.  At this ratio
at which, based on coherent state detection probability calculation,
the average photon number is $\mu = 1$. The power meter and the
id-201 were both employed for the initial calibration only, and
these two independent approaches agree each other within the error
bars in the DE measurement results. In subsequent measurements, we
use the attenuator only to set the photon flux.}

\begin{figure}
\includegraphics[width= 0.8 \textwidth]{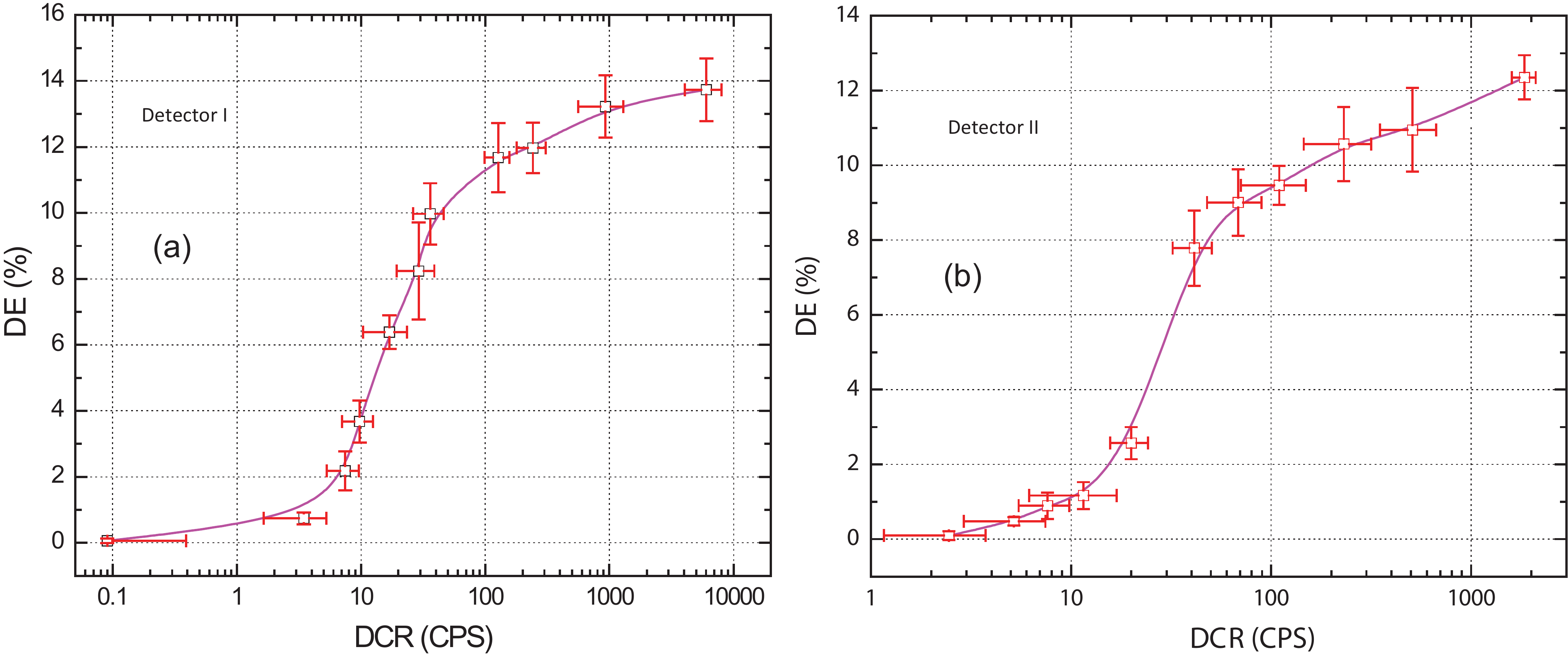}
\caption{DE vs Dark count measurements results for two NFAD
detectors. The discriminator threshold voltage level is set to 100
mV, and the testing temperature is 193 K.}\label{fig:Dark-DQE}
\end{figure}

The photon event, detected by the NFAD, is registered by a time tag
unit whose time resolution is 156 ps. The time window width was set
to be 1~ns for all of our measurements, unless explicitly mentioned.
Each measurement was carried out for a second and repeated 10 times.
Fig. \ref{fig:Dark-DQE} shows the DE measurement results using Eq.
\ref{eq:eta_approx}. The experimental DE exhibits a saturation at
darkcount rate above 10~kHz. The maximum detection efficiency for
the NFADs is $\sim 14\%$, and, with reasonable darkcount rates, it
can reach $\sim 10\%$.

\textcolor[rgb]{0,0.00,0.00}{The saturation  of detection efficiency
(at about $14~\%$, illustrated in Fig.~\ref{fig:Dark-DQE}) is
actually due to the run-away increase in apparent DCR by
afterpulsing. However, there is no way to separate the random
darkcount events from the afterpulses that they induce. The
afterpulsing process is highly non-linear, since afterpulse events
can induce additional afterpulsing events.  Hence, there is always a
point at which afterpulsing appears to increase quite dramatically
with any additional bias. Nonetheless, it is expected that NFADs can
provide significantly higher DE values ($>25\%$), similar to that of
the id-201 detector, when operated in the gated mode with a
sufficient hold-off time, although this imposes a reduction in
detector ``availability", that is undesirable for some experiments.}

Note that this detector efficiency is at the system level; no
normalization was used to take into account the device active area,
nor the insertion loss of the coupling path. In the DE test for both
NFADs, we also vary the discriminator voltage levels. No significant
DE improvements can be found by choosing different threshold values.
\begin{figure}
\includegraphics[width=0.8 \textwidth]{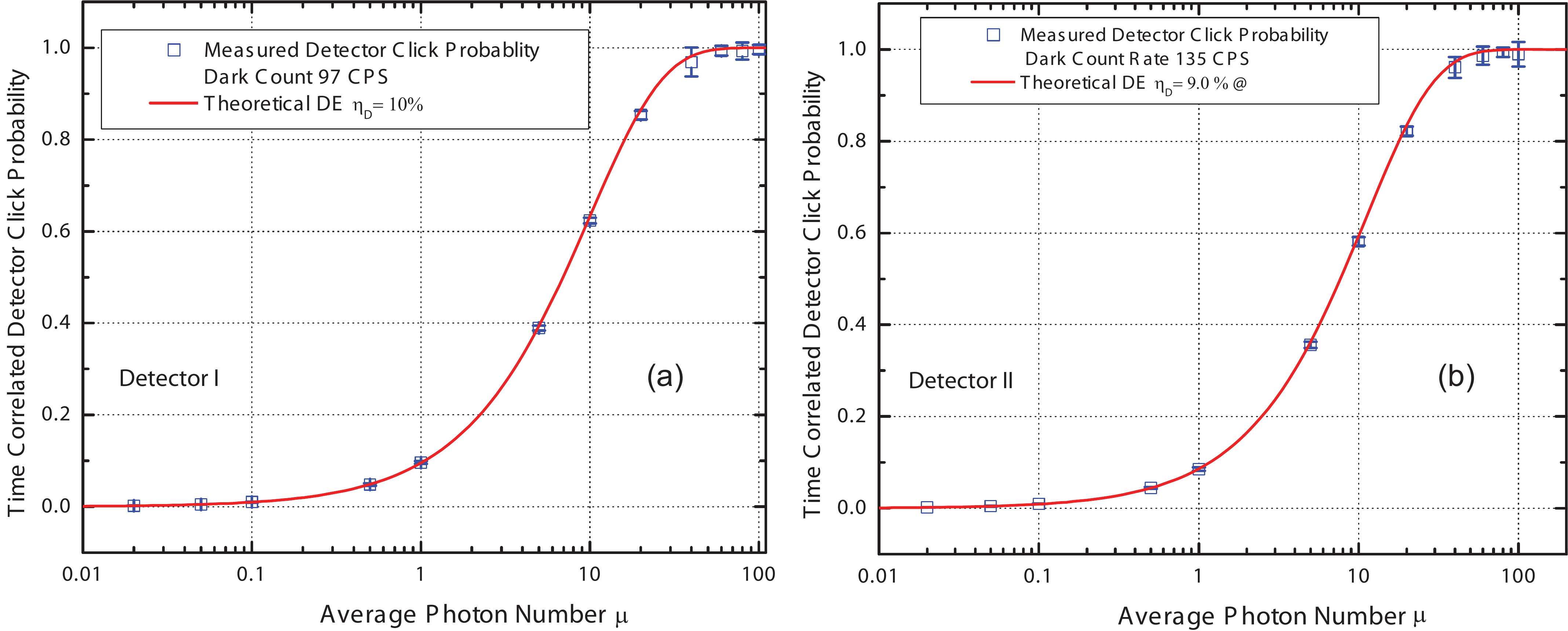}
\caption{Verification of bucket detector model. The hollow squares
are the experimental results.  The solid lines are the simulation
results based on modeling the detector as a bucket detector.
}\label{fig:DQE-PhotonNum}
\end{figure}

We verify that the detectors act as bucket detectors. If this
hypothesis is correct, then the detector click probability (the
ratio of the detection event counts within the correlation time
window to that of the total pulsed photon rate) is expected to
behave as $f(\mu) = 1- \exp(-\mathrm{\eta_D} \cdot \mu)$. The
experimental results are compared with theoretical curve in Fig.
\ref{fig:DQE-PhotonNum}. The curve is computed with the best fitting
value of DE, $\mathrm{\eta_D}$. The measurement results for two
different NFADs are obtained at the darkcount rate approximately at
100~CPS. The other testing conditions are kept the same the DE
measurement, except for the photon triggering rate, which is
increased to 10~kHz. The good agreement for both detectors indicate
that the bucket detector model is valid for our NFADs. This
description is thus appropriate for modeling further quantum optics
measurements.
\begin{figure}
\includegraphics[width= 0.8 \textwidth]{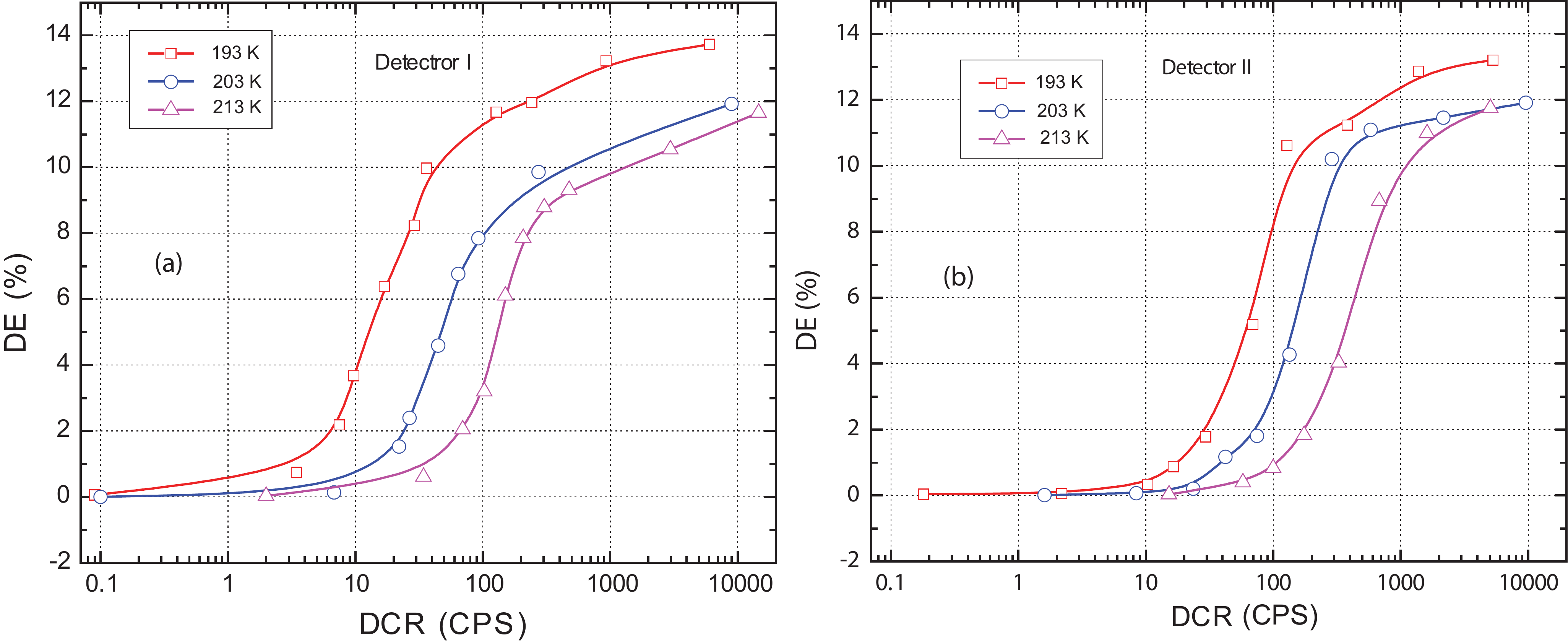}
\caption{Effect of temperature on detection efficiency and dark
counts. The data points represent experimental measurements, and the
lines are simple smooth fits.}\label{fig:DQE-Temp}
\end{figure}

A key benefit of our unique readout circuit is that we operate the
NFAD at lower temperature than most InGaAs SPADs are operated at
(typically around 233~K). Fig. \ref{fig:DQE-Temp} shows the effect
of temperature on the relation between DE and DCR. Once the DE and
its associated darkcount rate are measured, we can find the detector
noise equivalent power (NEP) via $\mathrm{NEP} = {{h\nu } \over
{\eta _\mathrm{D} }}\sqrt {2D}$. The best noise equivalent power is
estimated to be $8.1 \times 10^{-18}~\textrm{W Hz}^{-1/2}$ for
Detector~1; and $1.4 \times 10^{-17}~\textrm{W Hz}^{-1/2}$ for
Detector~2 at 193 K. We will also study the figure of merit, which
depends on the detector jitter time, in Section \ref{sec:Jitter}.
\subsection{DE  measurements using a continuous wave pumped SPDC source}\label{sec:SPDC}
\textcolor[rgb]{0.00,0.00,0.00}{Many quantum communication protocols
are based on entangled photon pairs created from SPDC. It is of
interest to characterize NFADs using photons from a SPDC source for
several reasons: the non-Poissonian photon distribution of SPDC
source behaves with a thermal-like property
\cite{Mollow_PhysRev_67_I}; Most importantly, correlated photons
allow to verify the detection efficiency with least assumption on
photon flux calibration \cite{Migdall_JMO_04}, and avoid such
calibration uncertainties in the WCP scheme; Moreover, the photon
pairs are created at random times, allowing us to convincingly
demonstrate the free-running mode of NFADs.}

The unique property of SPDC photon pairs has been exploited to
characterize single photon detectors ever since the seminal work by
Burnham \textit{et al.}\cite{Burnham:70}, other researchers have
published their approaches \cite{Migdall_JMO_04, Polyakov:07,
Kwiat:94} to reduce the uncertainties in determining the absolute
detector efficiencies. The experimental accuracy is limited by a
wide varieties of nonideal factors \cite{Migdall_JMO_04}. The
typical experimental configuration for characterizing a SPDC source
operated with a continuous-wave (CW) laser, uses two free running
single photon detectors, and time correlation photon counting
logics. Note that the production times of photon pairs are random in
this case.

An alternative configuration \cite{Hadfield_JAP_07}, which is widely
used particularly when the down converted photon pairs in the 1550
nm wavelength range, can be the use of of one free running detector,
in combination with a gated SPAD, which typically has  a higher
efficiency.
\begin{figure}
\includegraphics[width=0.8 \textwidth]{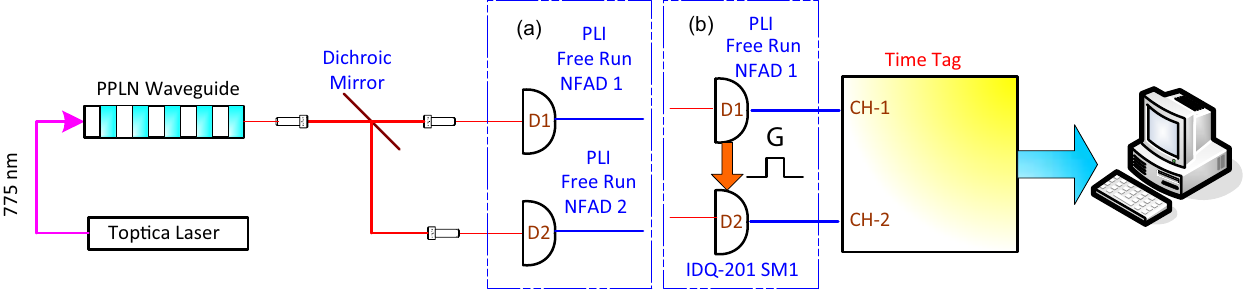}
\caption{Experimental setups used for two testing the NFAD~1 using
SPDC. In dashed box (a), two free running NFAD detectors. In dashed
box (b), one NFAD is placed in one of photon arms, a gated
commercial detector (IDQ) is at the other arm. Photon detections
from the NFAD are used to trigger an id-Quantique commercial single
photon detector.}\label{fig:SPDC_Testing_Scheme}
\end{figure}
\begin{figure}
\includegraphics[width= 0.8 \textwidth]{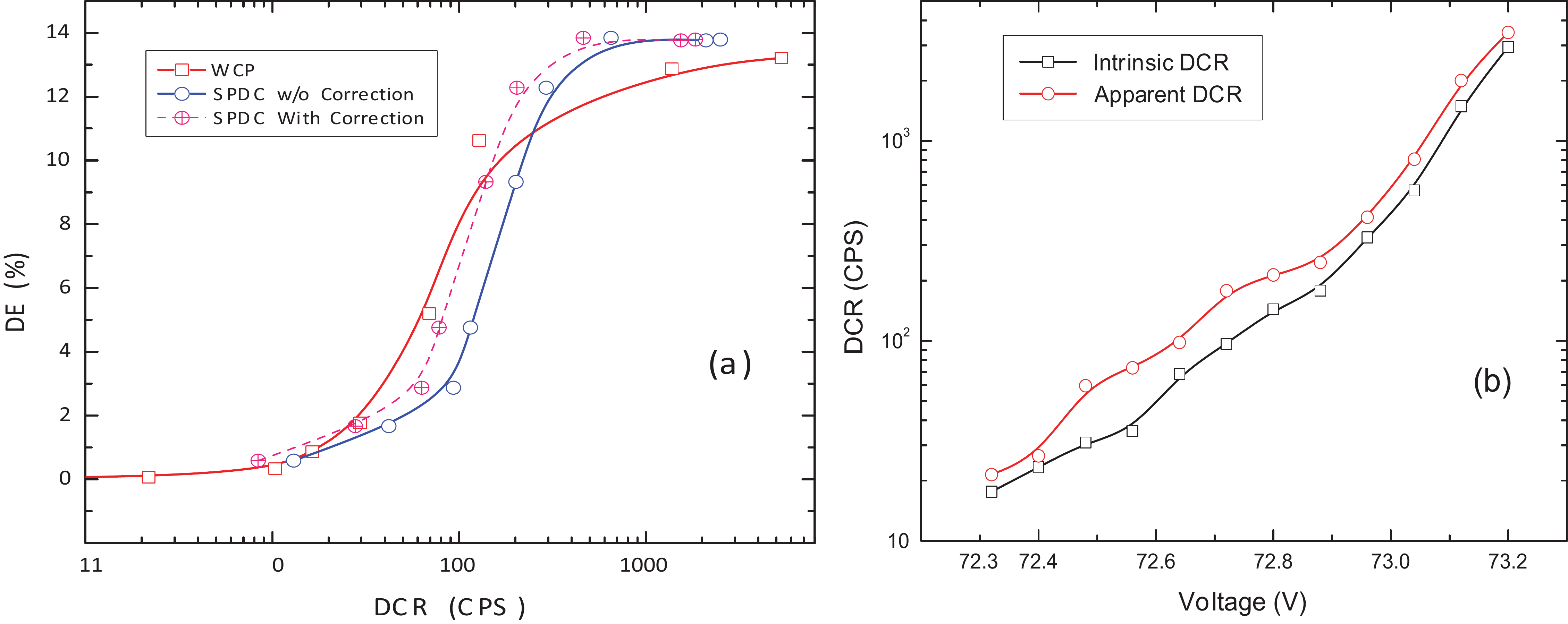}
\caption{(a) The inferred detection efficiency of the NFAD, from a
direct comparison with an id-201 commercial detector.
\textcolor[rgb]{0.00,0.00,0.00}{The solid line (hollow circle) is
the inferred DE using the SPDC scheme at 1530 nm; the dash line
(cross circle) is the inferred DE corrected for stray light
detection; the solid line (square) is the DE characterized by the
WCP scheme at 1550 nm. (b) The darkcount observed in the presence (
``apparent") and absence (``intrinsic") of stray light coupled into
the optical fiber connecting to the detector when varying the bias
voltage.}}\label{fig:SPDC_DE_Measurements}
\end{figure}

In our experiment, we tested  two combinations of detectors, as
displayed in Fig. \ref{fig:SPDC_Testing_Scheme}.  The SPDC source is
an integrated waveguide based on a periodically poled lithium
niobate (PPLN) substrate \cite{Tanzilli_PPLN_02}. Fig.
\ref{fig:SPDC_Testing_Scheme}~(a) illustrates the first setup, which
is comprised of two free running NFADs, with their detection events
recorded by a time tag unit. Both free-running NFADs have the figure
of merit $H > 1 \times 10^7$ at $\lambda =1550$ nm, as defined in
Ref. \onlinecite{Hadfield_NatPhoto_09}.

Fig. \ref{fig:SPDC_Testing_Scheme} (b) shows the  commonly used
configuration: a free running NFAD is employed to trigger the second
gated id-201 ULN SPD module.  Its settings are: detection efficiency
is $10 \%$, deadtime is 40 $\mu$s, and gate width is 2.5 ns. The
SPDC in PPLN is set to be in the non-degenerated mode, where the two
down converted photons have different wavelength of (1570 $\pm$
12.5) nm and (1530 $\pm$ 12.5) nm, respectively. The photons are
separated by a dichroic mirror and fed into two detectors connected
to channels 1 and 2.

The overall procedure is based on two successive measurements
depicted in Fig. \ref{fig:SPDC_Testing_Scheme} (a) and (b). For the
two free running detector scheme, we keep NFAD 1 at $6.1 \times
10^3$ counts per second on average. To achieve this condition, we
maintain a constant bias voltage to NFAD 1 and a constant pumping
power to the PPLN. We vary the bias voltage to the second detector
NFAD 2. At each voltage step, we record the total single counts, as
well as coincidental counts using the time tag unit.

The net count (total single rate minus darkcount rate) of the down
converted photons in channel one (CH-1) and  channel two (CH-2) are
given by $N_1 = \eta_1 \eta_{D1} N (1 + \kappa _1)$, $N_2 = \eta_2
\eta_{D2} N (1 + \kappa _2)$; their coincidental counts $N_c$ is
given by $N_c = \eta_1 \eta_{D1} \eta_2 \eta_{D2} N$, where $N$ is
number of photon pairs, $\eta_1$ and $\eta_2$ are the efficiencies
of the two channels, and $\eta_{D1}$ and $\eta_{D2}$ are the
detection efficiencies of the two detectors; $\kappa_1$ and
$\kappa_2$ are the afterpulsing coefficients of NFAD~1 and NFAD~2,
which will be addressed in the following section. Therefore, from
the scheme in Fig.~\ref{fig:SPDC_Testing_Scheme}~(a),  we can find
the ratio $ \chi_1 ={ {\eta _2 \eta _{D2}^\mathrm{NFAD} } \over {1 +
\kappa _1 }} = N_c / N_1 $ using two NFADs; Subsequently, if we
switch to the scheme in Fig.~\ref{fig:SPDC_Testing_Scheme}~(b), we
can find another ratio $\chi_2 ={ {\eta _2 \eta _{D2}^\mathrm{IDQ} }
\over {1 + \kappa _1 }} = N_c ^\prime / N_1^\prime $. Therefore, we
can infer the efficiency of NFAD 2 via $\eta_{D2}^\mathrm{NFAD} =
\eta_{D2}^\mathrm{IDQ} \chi_1 / \chi_2$. In practice, the
afterpulsing coefficient is hard to maintain constant when the
darkcount rate is high. Additionally, the accuracy of this detector
efficiency measurement is also limited by wide varieties of other
nonideal factors \cite{Migdall_JMO_04}.

Fig. \ref{fig:SPDC_DE_Measurements}~(a) displays the inferred
detection efficiencies $\eta_{D2}^\mathrm{NFAD}$ of NFAD~2, and the
results are compared with the aforementioned WCP method. The two
plots are at the same bias voltage and discriminator threshold
settings. The calculation of $\eta_{D2}^\mathrm{NFAD}$ is based on
the assumption that the channel~2 efficiency $\eta_2$, and NFAD~1
afterpulsing coefficient $\kappa_1$, are both kept constant in two
consecutive schemes. \textcolor[rgb]{0.00,0.00,0.00}{The DE inferred
from the SPDC method is plotted against apparent darkcounts when the
stray light photons are included. Those stray light photons impose
no effect on the inferred DE values, but reduce the SNR of the
NFADs. We performed a separate darkcount measurement to estimate the
number of darkcounts caused by stray light shown in
Fig.~\ref{fig:SPDC_DE_Measurements}~(b). This allows us to correct
the inferred DE values based on the substraction of stray light
detection. The corrected DE (the dash line in
Fig.~\ref{fig:SPDC_DE_Measurements}~(a)) exhibits improved SNR, and
the overall shape of the WCP and SPDC methods agrees.}
\section{Afterpulsing}
Afterpulsing is the most significant side effect for InGaAs SPADs;
it introduces extra counts, in addition to the intrinsic photon and
noise counts of the detector. It is mostly owing to defects in the
semiconductor materials form carrier traps which hold the avalanche
charge carriers. The lifetime of the trapped charge carriers ranges
from a few ns up to some tens of $\mu$s in conventional InGaAs/InP
SPADs working in the gated Geiger mode \cite{Tosi_JMO_09}. As the
operating temperature is lowered, such as the case in this work, the
afterpulsing problem becomes more prominent because the trapping is
longer.

\begin{figure}
\includegraphics[width=0.8 \textwidth]{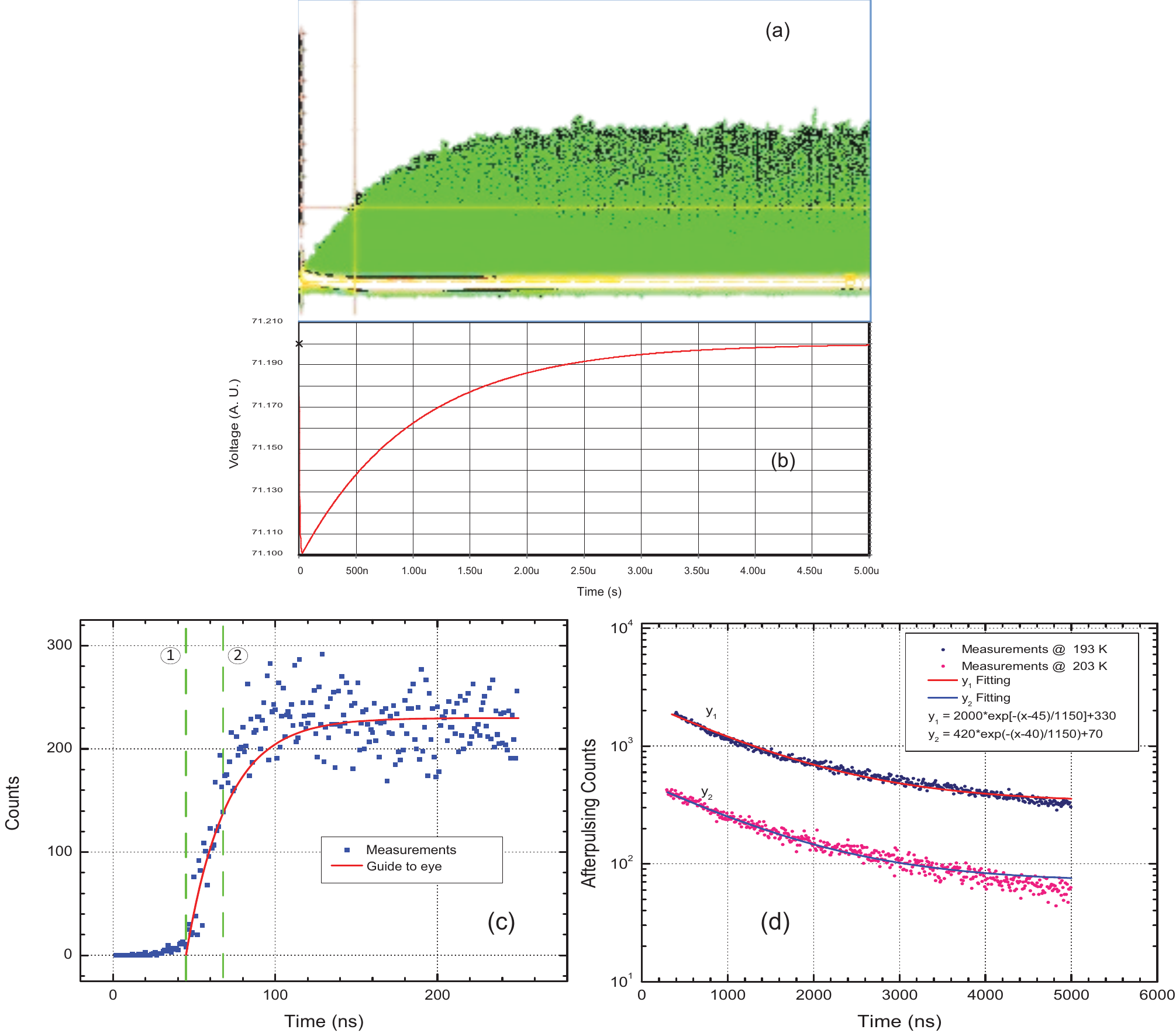}
\caption{(a)Accumulated measurements of the transient detector
response; this plot is the screen capture from a fast oscilloscope
collected for 6 hours; the saturated amplitude is about 300 mV. (b)
The simulation for the transient voltage drop across an intrinsic
SPAD inside the NFAD. Both plots share the same time scale of 5 $\mu
\textrm{s}$. (c) Measurement of the detector deadtime and recovery
time. The points are the measured photon detections in a 1~ns time
resolution, and the solid line is the guide to eye. (d) Afterpulsing
decay measurement and fitting. The random release of the trapped
charges induced by photon detection events results in the
afterpulsing probability decaying exponentially.  The simulation
curve is the solid line and the scattered dots are the accumulated
measurement results for two temperatures.
}\label{fig:NFAD_Voltage_transient}
\end{figure}

The NFADs afterpulsing measurements are performed with the same
setup as for the DE characterization (see Fig. 2). First, we measure
the photon response using a fast digital oscilloscope instead of the
time tagger. The measured screen shot is displayed in Fig.
\ref{fig:NFAD_Voltage_transient}~(a). We also simulate the voltage
across the intrinsic SPAD inside the NFAD with the SPICE circuit
model (refer to Fig. \ref{fig:SPICE_Model_Sim}~(b)). The transient
simulation results is shown in Fig.
\ref{fig:NFAD_Voltage_transient}~(b), where the time constant is
determined to be 1.1~$\mu\textrm{s}$ ($R=1.1~\textrm{M}\Omega$,
$C=1~\textrm{pF}$).

We use the time tagger  to measure the detector recovery. To
identify which event is the photon detection, we measure a
synchronization signal from the function generator on CH-1 (see Fig.
\ref{fig:Setup}) as the time reference. This enables us to
discriminate between photon detections and  events caused by dark
counts or afterpulsing. All of the afterpulsing measurements are
taken with the NFAD darkcount rate
\textcolor[rgb]{0.00,0.00,0.00}{\cite{Reply_6}} set to about 100~CPS
. Because we only record  time tags for detector click event where
the amplitude is higher than a  preset threshold voltage, the weak
avalanche events for the first few tens of ns after a detection are
not recorded.

Fig. \ref{fig:NFAD_Voltage_transient} (c) illustrates this effect as
an effective detector deadtime, following the rising voltage of NFAD
recovery process right after a photon detection event. Depending on
the operating temperature and the discriminator voltage settings, we
observed that the NFAD effective deadtime is at least 40 to 50 ns
long (dash line 1). After the deadtime period, we find a rapid
recovery of afterpulsing events,  rising within a time of about 25
ns (between dash line 1 and line 2).

In Fig. \ref{fig:NFAD_Voltage_transient} (d), we quantitatively
measure the lifetime of the trapped avalanche carriers using a
standard exponential decay function
\cite{Jiang_08_JQE_afterpulsing}, $C_0 \exp[-(t - T_\mathrm{dead}) /
\tau_\mathrm{d}]$. $T_\mathrm{dead}$, in our case, is the NFAD
effective deadtime resulting from the finite threshold of the time
tag unit. Theoretically, this value is around zero. With higher
temperatures, the effective deadtime is slightly shorter, as shown
in Fig. \ref{fig:NFAD_Voltage_transient}. $\tau_\mathrm{d}$ is the
characteristic detrapping time. Our best fit value to this curve is
$\tau_\mathrm{d} =$ 1.1 $\mu$s at both 193~K and 203~K, while the
prefactor $C_0$ is smaller at 203~K. If we compare $\tau_\mathrm{d}$
with voltage recovery time constant, it turns out that they are
almost identical, which implies that detrapping of the residual
avalanche charge is mainly induced by the detector recovery voltage.

\begin{figure}
\includegraphics[width=0.8 \textwidth]{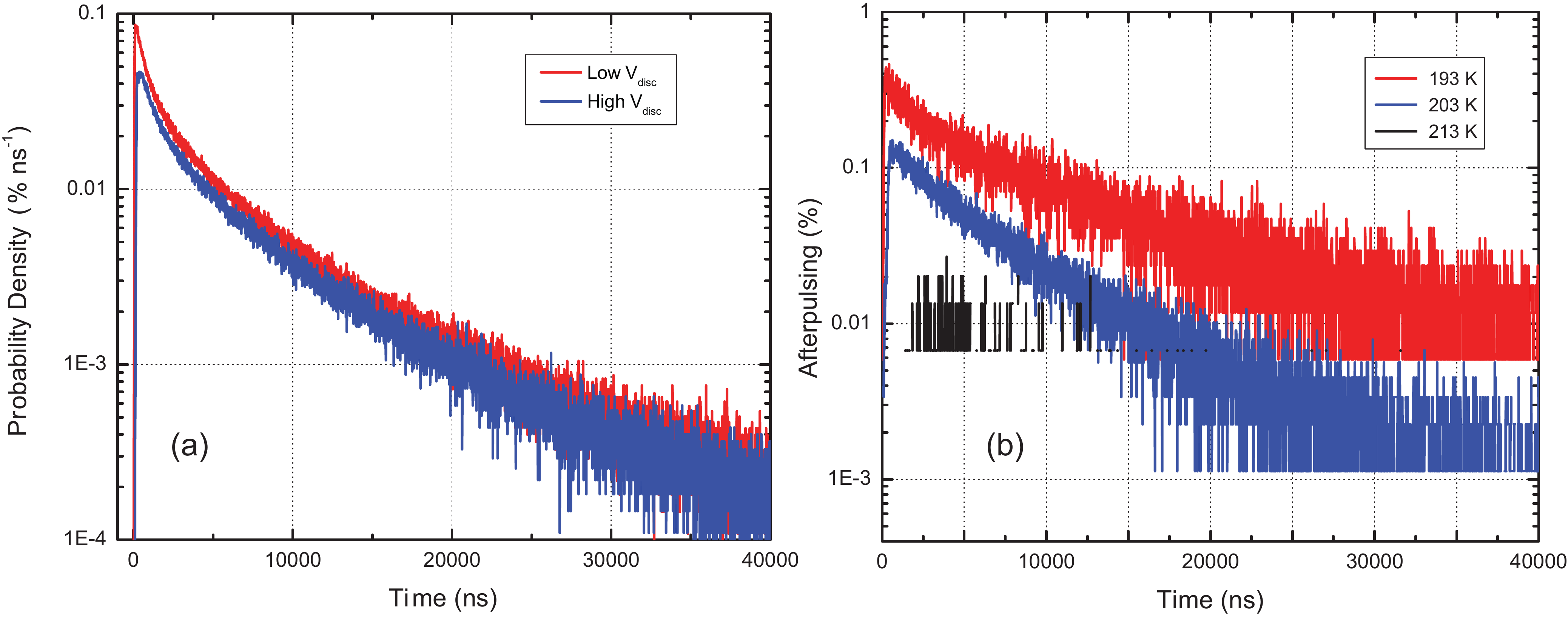}
\caption{(a) Afterpulsing under two threshold levels of the time tag
unit input.  The temperature is at 193~K. (b) Afterpulsing
probability density at three distinct temperatures, while the
darkcount rate was kept constant at about 100 CPS. All data are
collected for 1000 s. The threshold was set to be 0.1 V. The
reference clock rate, used to synchronize the time frame, is set to
10 kHz.}\label{fig:Afp_Vths}
\end{figure}
Fig. \ref{fig:Afp_Vths} (a) shows the afterpulsing behavior of our
NFADs for different discriminator voltages.  We notice that
increasing the threshold voltage can reduce the amount of
afterpulsing for a short time after the detection, but the effect on
the overall number of afterpulsing events is small. Fig.
\ref{fig:Afp_Vths} (b) illustrates the effect of operating
temperature on afterpulsing probability. As expected, afterpulsing
becomes more prominent at lower temperatures.  However, it is
important to note that the measurements are performed at constant
darkcount levels, and therefore the detector efficiency is lower for
the higher temperature settings.

\begin{figure}
  \includegraphics[width= 0.8 \textwidth]{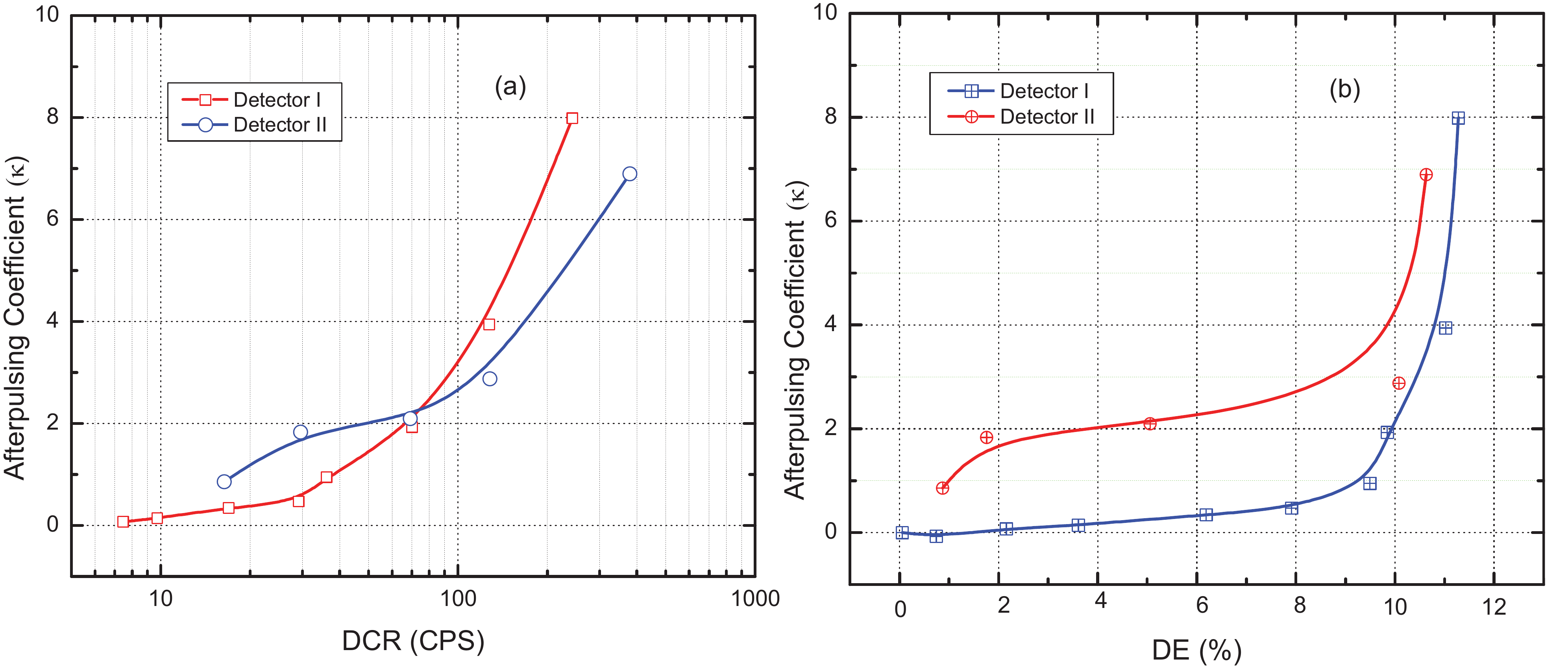}
\caption{Afterpulsing parameter, $\kappa$, measurement at 193~K by
WCP characterization setup for both two NFADs. (a) The afterpulsing
coefficient as a function of the darkcount rate; (b) The
afterpulsing coefficient as a function of the detection efficiency.
The wavelength of probing WCP pulses is 1550 nm for all
measurements.}\label{fig:AfterpulsingKappa}
\end{figure}

Another interesting approach to model afterpulsing behavior of a
NFAD is by introducing afterpulsing coefficient defined as $ \kappa
= {{N_t - N_c  - D} \over {N_c }}$, where $N_c$ is the correlated
count rate, $N_t$ is the total raw count rate, and $D$ is the
darkcount rate at a given bias voltage. We vary the bias voltage to
collect the raw counts and compute $\kappa$. The results are
displayed in Fig. \ref{fig:AfterpulsingKappa} (a) and (b).

\textcolor[rgb]{0.00,0.00,0.00}{The comparison of the afterpulsing
performance of two detectors indicates that Detector I has a better
afterpulsing performance than that of Detector II. We attribute this
to two factors: (1) Detector II has a larger active area than
Detector I. Consequently, there are more defect trap sites in the
active region that give rise to afterpulsing effects, by trapping
carriers from one avalanche and releasing them at a later time when
the detector becomes re-armed. (2) The larger area has a larger
capacitance. This results in a proportionally larger number of
charges generated when the capacitance is discharged with each
avalanche.  This increased charge flow per avalanche gives rise to
more trapped carriers at defect sites, and thus the afterpulsing
effect increases since more carriers are trapped and subsequently
detrapped. The combination of these two effects creates a
significant rise in afterpulsing events as the detector size is
increased. The unwanted effects can be alleviated using active
hold-off circuitry, which is widely used in gated mode detectors.
However, we chose to keep the circuit as simple as possible, in
order to avoid complexity of the electronics, its addition of stray
capacitance, and undesirable elevate detector temperature.}

Note that the coefficient $\kappa$ is obtained via the WCP scheme.
As we mentioned in the previous section, due to changes in
experimental conditions including surround light sources, the
observed darkcount rates in the SPDC experiments were different than
for the WCP scheme. To estimate $\kappa$ using darkcount rate for
the SPDC scheme, one must take into account the joint detection
efficiency.
\section{Time Resolution}\label{sec:Jitter}
For quantum optics applications, the timing jitter of the detector,
$\tau _\mathrm{jitter}$, is another key parameter. The jitter time
of the detector results from the fluctuations of its response time.
Many mechanisms contribute to this variation. One source of jitter
is the readout electronics.  Another is intrinsic to the detector,
primarily due to the stochastic build-up time of avalanche carriers
within the multiplication region \cite{Jiang_SPIE_11_NFAD} when the
SPD detects a photon.

Here, we measured the jitter time using a Becker $\&$ Hickl SPC-130
PCI card. We change the gain of the time to amplitude converter
(TAC) to set the time resolution to 1.22 ps per division. This
setting is used for all of the measurements. The other key technique
for measuring high precision jitter time is to carefully adjust the
zero crossing (ZC) level of both inputs, usually set close the
threshold levels.
\begin{table}[h]
  \centering
\caption{Timing jitter measurement \cite{Reply_8a} at 10 KHz
repetition rate of pulse laser, 0.1 photon and 1.0 photon on average
per pulse. The measurements are accumulated for 1000
s.}\label{table_1}
  \begin{tabular}{c|c|c|c|c|c|c}
    \hline
    \hline
    Low Limit & ZC Level & Total Count & Peak Count & FWHW  & Dark Count & Average \\
    (mV)    & (mV)  & & & (ps) & (CPS) & Photon number\\
    \hline
       -150 & -86 & 100261 & 799 & 154 & $\sim$ 150 & 0.1 \\
    -150 & -96 & 95620 & 1186 & 92 & $\sim$ 150 & 0.1 \\
    -140 & -96 & 96172 & 1163 & 69 & $\sim$ 200  & 0.1 \\
    -110 & -96 & 88066 & 1361 & 70 & $\sim$ 150 & 0.1 \\
     -100 & -96 & 88118 & 1405 & 59 & $\sim$ 150 & 0.1 \\
    -90 & -96 & - & - & - & - & 0.1\\
    \hline
    -150 & -96 & 706596 & 15780 & 31 & $\sim$ 150 & 1.0 \\
    \hline
  \end{tabular}
\end{table}
\begin{figure}
\centering
\includegraphics[width= 0.6 \textwidth]{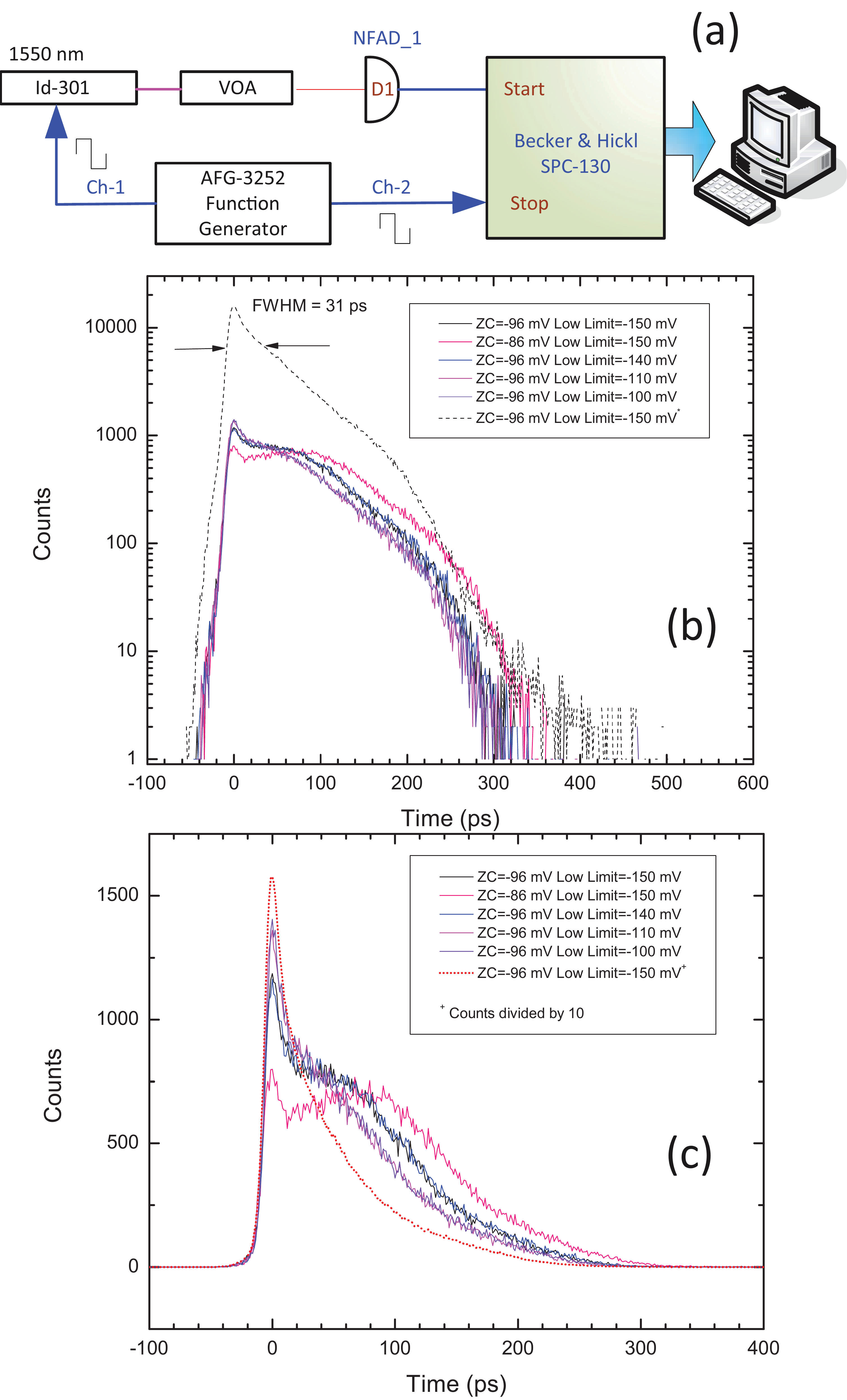}
\caption{Measurement of the detector jitter
\textcolor[rgb]{0.00,0.00,0.00}{by Becker $\&$ Hickl SPC-130
time-correlation card with a time resolution of 1.22 ps.} (a)
Experimental setup; (b) Measurements plots in logarithmic scale; (c)
Normalized measurement plots in linear scale. The data are taken
with 0.1 (sold lines) and 1.0 (dash line) photons per pulse for 1000
s, with different combinations of settings to the constant-fraction
discriminator (CFD), \textcolor[rgb]{0.00,0.00,0.00}{the start
signal for the time measurement.} The function generator (AFG~3252),
which produces the stop signal, is connected to the ``Synch" input
of the card. The detector output is connected to CFD input after an
pulse inverter transformer. \textcolor[rgb]{0.00,0.00,0.00}{The
average photon number per pulse, from 0.1 to 1, is controlled by
changing the variable attenuator (VOA) settings}.
}\label{fig:Jitter01}
\end{figure}

\textcolor[rgb]{0.00,0.00,0.00}{The measurement setup is illustrated
in Fig. \ref{fig:Jitter01} (a).  The results are plotted in both
logarithmic and linear scales by Fig. \ref{fig:Jitter01} (b) and
(c), respectively.} We took measurements for two different values of
the average photon number per pulse. The low average photon number,
0.1 photon per pulse, effectively simulates the regime of single
photon detection. At this level of $\mu$, the jitter time is mostly
below 150 ps, which is compatible to the resolution of our time tag
unit. This photon flux level simulates the case present in many
quantum optics experiments, such as detecting photon pairs generated
by nonlinear optical processes. We also set $\mu=1$, not only a
comparable photon flux to what is used for decoy state QKD protocol
\cite{Lo:PhysRevLett_Decoy05}, but also to get an indication of the
timing jitter intrinsic to the electronics. In all of the tests, we
maintained the darkcount rate below 200~CPS.

The measurement settings and jitter measurement are also summarized
in Table \ref{table_1}.  For $\mu=1$, the measured jitter is 31 ps.
For $\mu=0.1$, the timing jitter becomes wider also due to the width
of laser pulse\cite{Reply_8a}, which is specified to be $<$300~ps.
Most measured jitters are around 100 ps, depending on the setting of
PC-130 card ranging, ranging from 59 ps to 154 ps. Most importantly,
all of the detection events fall within a 400 ps time length,
regardless of measurement settings, which
\textcolor[rgb]{0.00,0.00,0.00}{by itself} is a very good timing
response. The typical response of SPADs, including silicon ones, is
that their time distribution has a long tail on the order of a few
ns (see Ref. \onlinecite{Hadfield_NatPhoto_09}) even if their
FHWM-jitter is a few tens of ps.

Table \ref{table_1} also suggests a trade off between the total
counts and the combination of ``Low limit" and ``ZC level" settings,
by which we can collect higher total counts, but at the cost of
longer jitter time. However, no long recovery tail is found in all
of our jitter measurements. Lastly, based on these jitter
measurements, the detector figure of
merit\cite{Hadfield_NatPhoto_09} can be calculated from the
expression $ H = {{\eta _D } \over {D \cdot \tau _{jitter} }}$. Our
observations show that the best figure of merit is  $6.3 \times
10^7$ for Detector 1.
\section{Showing feasibility of entangled photon QKD over 400 km telecom fiber}
\begin{figure}[h]
\includegraphics[width= 0.8 \textwidth]{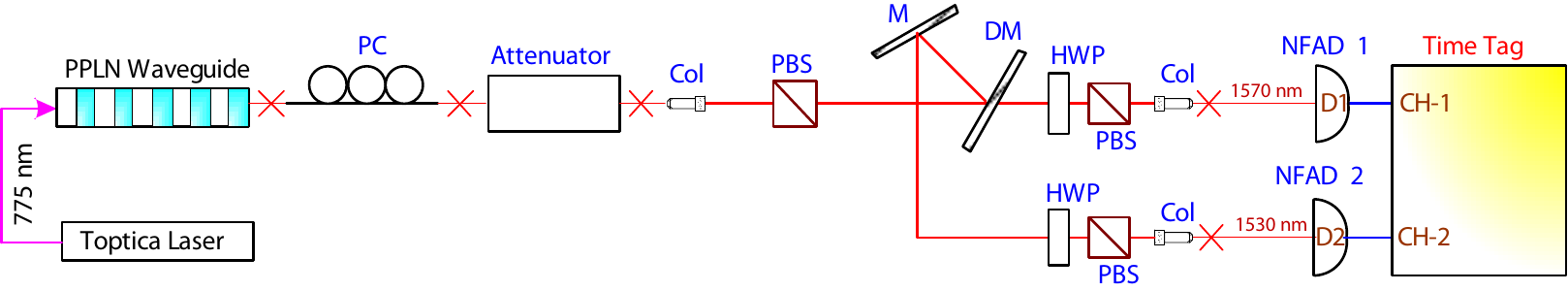}
\caption{Experimental setup for the investigation of High Loss QKD.
Photon pairs are produced in a PPLN waveguide, and are sent through
a variable attenuator. They are then split up by a dichroic mirror
(DM) and detected by the NFADs.  PBS: polarizing beam splitter;
Attenuator: programmable optical attenuator; HWP: half wave plate;
M: mirror; Col.: collimator.}\label{fig:HiLossSPDC_Setup}
\end{figure}

Entanglement-based QKD can achieve the longest distance when the
photon pair source is symmetrically put in the middle of two parties
Alice and Bob \cite{Scherer_PRA_09, Scherer_OptExpress_11}. Ma
\textit{et al.} \cite{Ma_PRA_07} first showed this configuration can
tolerate up to 70~dB channel loss. Later Scheidl \textit{et al.}
\cite{Scheidl_NJP_09} performed the experiment and verified that the
highest loss of their entanglement QKD achieved 71~dB. Here, we used
a similar source, i.e. a CW pumped SPDC photon pairs. The quantum
channel loss is simulated using a programable optical attenuator.
The experimental setup is illustrated by Fig.
\ref{fig:HiLossSPDC_Setup}.  To demonstrate the feasibility of a
real QKD system, our setup has the same optical components as a
standard QKD receiver, therefore we consider all potential losses
from Alice, Bob and the entangled photon source.

\begin{figure}
\centering
\includegraphics[width=0.8 \textwidth]{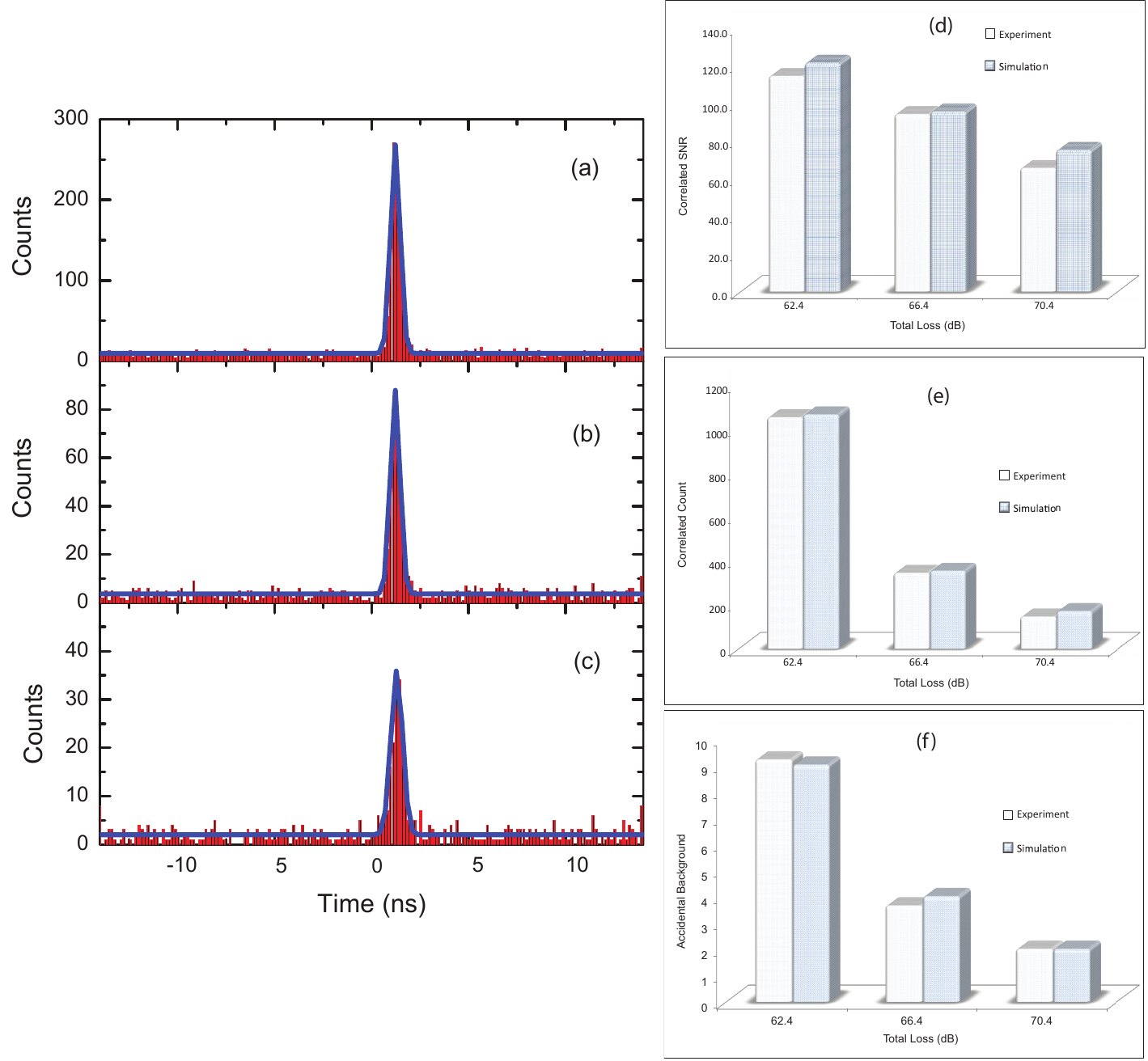}
\caption{The accumulated counting results
\textcolor[rgb]{0.00,0.00,0.00}{collected by a 16-channel time tag
unit with a time resolution of 156 ps, at }three levels of
attenuation settings, for (a) 62.4 dB and 24 hours collection time;
(b) 66.4 dB and 20 hours collection time; (c) 70.4 dB and 24 hours
collection time. Solid blue lines are Gaussian distribution
fittings, with each of them having $\sigma$=0.25 ns ,
\textcolor[rgb]{0.00,0.00,0.00}{which is dominated by the timing
resolution of the time tag unit}. (d)-(f) are the experimental
results (left) in comparison with the simulation results (right). In
each of them, all above three levels of channel loss are taken into
account. (d) accidentals-SNR: the ratio between the correlated and
accidental counts; (e) the total correlated counts; (f) the total
accidental background counts.}\label{fig:HiLossSPDC_CountingReults}
\end{figure}

We performed 68 hours of measurements to test three attenuation
values set by the variable optical attenuator. We use a time tag
unit to collect the counts from two NFAD detectors. The time
resolution can be chosen in the postprocessing stage. The internal
optical losses are characterized for two wavelengths, averaging 8.22
dB per arm. The total loss is determined to be 62.4 dB, 66.4 dB and
70.4 dB by separate measurements. The darkcount rate for both
detectors is set to be below 100 CPS. Fig.
\ref{fig:HiLossSPDC_CountingReults} displays the summary of our
experimental results in (a)-(c), as well as measurements in
comparison with to the simulated results in (d)-(f). In the
simulation, the detector efficiencies are chosen to be $\eta_{D1} =
8 \%$ at 1570 nm wavelength and $\eta_{D2} = 6 \%$ at 1530 nm
wavelength.

The analysis of the experimental data and simulation of secret key
rate is based on the framework of Ref. \onlinecite{Scheidl_NJP_09}.
However, we only focus on the accidentals visibility by looking at
the ratio between the correlated and accidental counts.
\textcolor[rgb]{0.00,0.00,0.00}{We also assume our source is an SPDC
source with its characteristic emission statistics, operating at the
average photon pair rate $N$}, is very low (4.5 MHz in our
experimental measurements). $\eta _A$ and $\eta _B$ are the channel
total efficiencies, each of them contains four contributions: (1)
the finite detector efficiency ($\eta_{D1}$, $\eta_{D2}$); (2) PPLN
source coupling loss; (3) receiver optical loss ($\eta_{CA}$,
$\eta_{CB}$) and (4) link attenuation loss ($t_{SA}$, $t_{SB}$),
which is set by the programable attenuator. So $\eta _A = \eta_{CA}
t_{SA} \eta_{D1}$, and $\eta _B = \eta_{CB} t_{SB} \eta_{D2}$. In
general, $\eta _A \ll 1$ and $\eta _B \ll 1$ due to photon
transmission losses. Based on these conditions, the lower bound
 of quantum bit error rate\cite{Reply_Eq3} $Q_{BER}$ is:
\begin{equation}\label{eq:QBER}
Q_{BER}  = \left\{ {2 + {{\eta _A \eta _B } \over {\mu _P \left[
{\left( {1 + \kappa _A } \right)\eta _A  + {{D_A } \over N}}
\right]\left[ {\left( {1 + \kappa _B } \right)\eta _B  + {{D_B }
\over N}} \right]}}} \right\}^{ - 1}.
\end{equation}
where the average photon pair number per time window $\mu _P  = NW$.
Here, $W$ is the time window, and $N$ is the average photon pair
rate. It is limited by the detector jitter time as well as the
resolution of the time tag unit. From the previous experiments, we
determined that jitter time of the NFADs is shorter than 150~ps,
which is shorter than the time resolution of the time tag unit
156~ps \textcolor[rgb]{0.00,0.00,0.00}{(from DotFast / UQDevices
\cite{dotfast}). Note that the Becker $\&$ Hickl card, which  has
much better time resolution, could not be used in this application
because it is not suitable for the continuous time stamping of
photon pair detections over the long accumulation times we require
($\geq~20$ hours). Consequently, in all of our measurements and
simulation, the smallest time window is chosen to be 156 ps. Based
on this timing resolution, $N$ could be as high as 1 GHz and still
provide a 9:1 ratio for the 1-photon pair probability versus the
2-photon pair probability. } We assume $\kappa_A=\kappa_B=2$ as the
afterpulsing coefficients to fit both detectors (measurements shown
in Fig.~\ref{fig:AfterpulsingKappa}), given the detector
efficiencies in conjunction with their darkcount settings. $D_A$ and
$D_B$ are the darkcount rates, and are both set to 100 CPS.

Using the QBER obtained from Eq.~\ref{eq:QBER}, we can estimate the
secret key rate, $R$, by:
\begin{equation}\label{eq:SecureKeyRate}
R \ge {1 \over 2}\left\{ {N\eta _A \eta _B \left[ {1 - f\left(
{Q_{BER} } \right)H_2 \left( {Q_{BER} } \right) - H_2 \left(
{Q_{BER} } \right)} \right]} \right\},
\end{equation}
where $H_2$ is the binary entropy function, $H_2 \left( x \right) =
- x\log _2 \left( x \right) - \left( {1 - x} \right)\log _2 \left(
{1 - x} \right) $. The correction factor is $f\left( {Q_{BER} }
\right) = 1.22 $.
\begin{figure}
\includegraphics[width= 0.8 \textwidth]{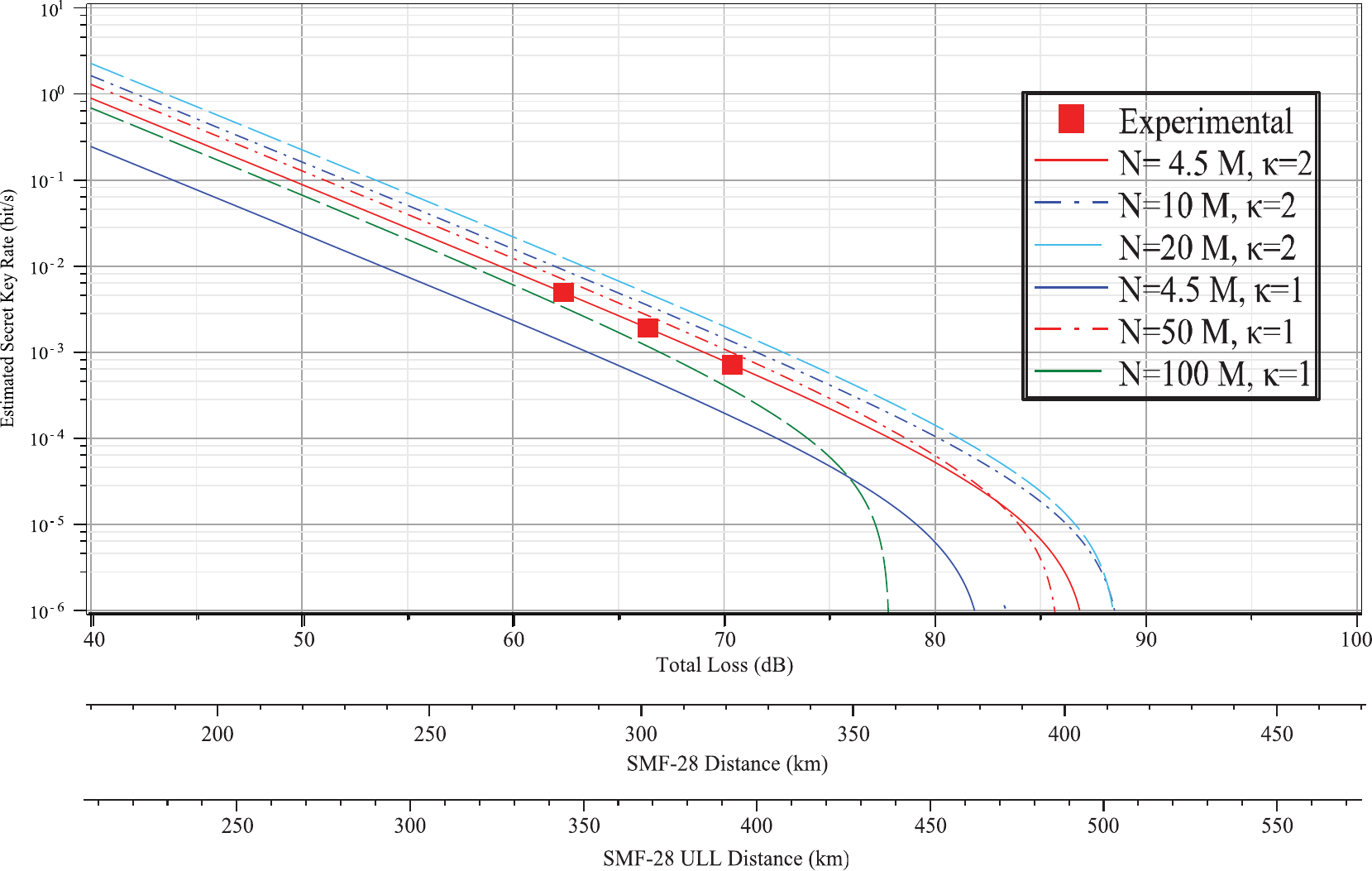}
\caption{ The estimated secret key rate for entanglement QKD versus
the total channel loss. \textcolor[rgb]{0.00,0.00,0.00}{In the
numerical simulation: for the cases of $\kappa=2$, we use detector
efficiency of $8\%$ and $6\%$, respectively; and for the cases of
$\kappa=1$, two detector efficiencies are chosen to be $4\%$ and
$3\%$, respectively.} The solid squares are based on actual
measurements of the signal to noise for a given loss. The lines show
secret key rates calculated for different source rates and
afterpulsing coefficient. Two additional x-axes show corresponding
QKD distances according to the total channel loss, for standard
SMF-28 single mode fiber and ultra low loss fiber, SMF-28 ULL. These
results indicate that entanglement-based QKD over 400 km is
feasible. }\label{fig:HiLossQKD_Distance}
\end{figure}

We experimentally simulated secret key rates for two types of
telecom optical fibers: a standard telecom fiber (SMF-28) whose
loss, $\alpha_L$, is 0.2 dB per km, and an ultra low loss (ULL)
telecom fiber (SMF-28-ULL) \cite{Stucki_NJP_09}, whose loss is 0.164
dB per km. Note that in both cases, we have introduced 6 dB extra
loss, to account for optical losses including interconnections at
Alice, Bob and central source sites. The results are showed in Fig.
\ref{fig:HiLossQKD_Distance}. The discrete dots in the figure are
based on the experimental data for three levels of optical
attenuations. The results suggest that the distance of entangled
pair based QKD is feasible over 300 km via the standard telecom
fiber, and can reach over 400 km in telecom ULL fiber. One caveat is
that the finite key size effect has not been taken into account for
both simulations. Nonetheless, given the time resolution of our
detector and the possibility to achieve an entanglement photon
source whose rate can reach up to a few hundreds of MHz,
entanglement-based QKD system over 400 km is possible.

Another remarkable observation is that normally a detector free of
afterpulsing ($\kappa = 0$) will have its cut-off QKD distance limit
proportional to the photon pair rate. This is not the case for the
current NFAD design, because our detector is free running and no
extra afterpulse suppressing circuit has been introduced, leading to
its afterpulsing coefficient of $\kappa = 2$. With this nonzero
value, the QKD cut-off distance is no longer proportional to $N$.
Instead the secret key rate vanishes when $N > 70$ MHz
approximately. However, one remedy to this situation is to decrease
the current detector efficiency to $4 \%$ and $3 \%$, respectively.
As a result, the afterpulsing parameter drops to less than one.
Consequently, in the case of $\kappa = 1$, the secret key rate does
exist even when $N =$ 100 MHz. However, as indicated in Fig.
\ref{fig:HiLossQKD_Distance}, its secret key rate is lower than the
case when $N =$ 20 MHz, and the efficiency of both detectors are
twice as high with $\kappa = 2$. Nevertheless, the regime where
$\kappa = 1$ will be beneficial for studying novel high rate SPDC
entanglement photon pair source.
\section{Conclusion}
We have reported a novel design to read out single photon responses
for NFAD operating at low temperatures. The system circuit diagram
has been presented in detail. We carried out a SPICE model
simulation to compare the conventional readout method with our
approach. The simulation result agrees with the experimental
measurement. Detection efficiencies of both NFADs are characterized
using TCSPC measurements under different temperatures and threshold
settings via WCP source. Subsequently, we have used two free running
NFADs and compared them with an id-201 detector. The inferred DE of
NFAD~II is derived from coincidence counts of the id-201 matches the
expected NFAD efficiencies obtained in WCP characterization. By
scanning the average photon number in WCP, we confirmed that the
NFAD acts as a bucket detector \cite{Jennewein_JMO_11} in the
quantum optics model. We also found that the detection repetition
rate can be close to 100 kHz without DE degradation. The
afterpulsing was tested by varying the threshold settings and
ambient temperatures. The temporal distribution of afterpulsing
events follows an exponentially decaying trend. Our measurement of
the NFAD jitter time can achieve a FWHM of 31 ps. The dark noise
below the rate of $\mathrm{10^{-7} ns^{-1}}$ for the best NEP $8.1
\times 10^{-18}~\mathrm{W \cdot Hz^{-1/2}}$. The figure of merit for
SPDs was determined to reach $6.3 \times 10^7$. We also demonstrated
that an entanglement-based QKD system can tolerate higher losses up
to 70 dB, which infers successful key distribution distance over 400
km of ULL telecom optical fiber.

We like to stress that the operating temperature for NFAD is
maintained in continuously cryogenic free mode with a simple
electronic circuitry. No cryogenic facility is involved; no vacuum
is needed. The cooling power of our system ($10~\textrm{W})$ allows
it to be an readily expandable platform to include higher numbers of
NFADs, where the heat emitted from each detector is less than 1 mW,
even for count rates of about 1~MHz detection pulse rate per
detector.  These advantages enable this NFAD system to be a good
candidate in quantum optics, quantum cryptography, and SPDC
entanglement source characterization applications for the telecom
band.
\section{Acknowledgements}
This work is financially supported by the Natural Sciences and
Engineering Research Council of Canada (NSERC), Industry Canada;
Canada Foundation for Innovation (CFI), Canadian Institute for
Advanced Research (CIFAR), and Early Researcher Awards (ERA)/MEDI
Ontario. We are grateful for useful discussions with Dr. L. K. Shalm
and Dr. H. H\"{u}bel. We wish to thank Zhenwen Wang for the
technical assistance.

\begin{thebibliography}{50}%
\makeatletter
\providecommand \@ifxundefined [1]{%
 \@ifx{#1\undefined}
}%
\providecommand \@ifnum [1]{%
 \ifnum #1\expandafter \@firstoftwo
 \else \expandafter \@secondoftwo
 \fi
}%
\providecommand \@ifx [1]{%
 \ifx #1\expandafter \@firstoftwo
 \else \expandafter \@secondoftwo
 \fi
}%
\providecommand \natexlab [1]{#1}%
\providecommand \enquote  [1]{``#1''}%
\providecommand \bibnamefont  [1]{#1}%
\providecommand \bibfnamefont [1]{#1}%
\providecommand \citenamefont [1]{#1}%
\providecommand \href@noop [0]{\@secondoftwo}%
\providecommand \href [0]{\begingroup \@sanitize@url \@href}%
\providecommand \@href[1]{\@@startlink{#1}\@@href}%
\providecommand \@@href[1]{\endgroup#1\@@endlink}%
\providecommand \@sanitize@url [0]{\catcode `\\12\catcode
`\$12\catcode
  `\&12\catcode `\#12\catcode `\^12\catcode `\_12\catcode `\%12\relax}%
\providecommand \@@startlink[1]{}%
\providecommand \@@endlink[0]{}%
\providecommand \url  [0]{\begingroup\@sanitize@url \@url }%
\providecommand \@url [1]{\endgroup\@href {#1}{\urlprefix }}%
\providecommand \urlprefix  [0]{URL }%
\providecommand \Eprint [0]{\href }%
\providecommand \doibase [0]{http://dx.doi.org/}%
\providecommand \selectlanguage [0]{\@gobble}%
\providecommand \bibinfo  [0]{\@secondoftwo}%
\providecommand \bibfield  [0]{\@secondoftwo}%
\providecommand \translation [1]{[#1]}%
\providecommand \BibitemOpen [0]{}%
\providecommand \bibitemStop [0]{}%
\providecommand \bibitemNoStop [0]{.\EOS\space}%
\providecommand \EOS [0]{\spacefactor3000\relax}%
\providecommand \BibitemShut  [1]{\csname bibitem#1\endcsname}%
\let\auto@bib@innerbib\@empty
\bibitem [{\citenamefont {Jiang}\ \emph
  {et~al.}(2009{\natexlab{a}})\citenamefont {Jiang}, \citenamefont {Itzler},
  \citenamefont {Nyman},\ and\ \citenamefont
  {Slomkowski}}]{Jiang_SPIE_09_NFAD}%
  \BibitemOpen
  \bibfield  {author} {\bibinfo {author} {\bibfnamefont {X.}~\bibnamefont
  {Jiang}}, \bibinfo {author} {\bibfnamefont {M.~A.}\ \bibnamefont {Itzler}},
  \bibinfo {author} {\bibfnamefont {B.}~\bibnamefont {Nyman}}, \ and\ \bibinfo
  {author} {\bibfnamefont {K.}~\bibnamefont {Slomkowski}},\ }in\ \href@noop {}
  {\emph {\bibinfo {booktitle} {Advanced Photon Counting Techniques III}}},\
  Vol.\ \bibinfo {volume} {7320}\ (\bibinfo  {publisher} {SPIE},\ \bibinfo
  {year} {2009})\ p.\ \bibinfo {pages} {732011 (10 pp.)}\BibitemShut {NoStop}%
\bibitem [{\citenamefont {Jiang}\ \emph {et~al.}(2011)\citenamefont {Jiang},
  \citenamefont {Itzler}, \citenamefont {O'Donnell}, \citenamefont
  {Entwistle},\ and\ \citenamefont {Slomkowski}}]{Jiang_SPIE_11_NFAD}%
  \BibitemOpen
  \bibfield  {author} {\bibinfo {author} {\bibfnamefont {X.}~\bibnamefont
  {Jiang}}, \bibinfo {author} {\bibfnamefont {M.~A.}\ \bibnamefont {Itzler}},
  \bibinfo {author} {\bibfnamefont {K.}~\bibnamefont {O'Donnell}}, \bibinfo
  {author} {\bibfnamefont {M.}~\bibnamefont {Entwistle}}, \ and\ \bibinfo
  {author} {\bibfnamefont {K.}~\bibnamefont {Slomkowski}},\ }in\ \href@noop {}
  {\emph {\bibinfo {booktitle} {Advanced Photon Counting Techniques V}}},\
  Vol.\ \bibinfo {volume} {8033}\ (\bibinfo  {publisher} {SPIE},\ \bibinfo
  {year} {2011})\ p.\ \bibinfo {pages} {80330K (9 pp.)}\BibitemShut {NoStop}%
\bibitem [{\citenamefont {Itzler}\ \emph {et~al.}(2010)\citenamefont {Itzler},
  \citenamefont {Jiang}, \citenamefont {Onat},\ and\ \citenamefont
  {Slomkowski}}]{Itzler_SPIE_10}%
  \BibitemOpen
  \bibfield  {author} {\bibinfo {author} {\bibfnamefont {M.~A.}\ \bibnamefont
  {Itzler}}, \bibinfo {author} {\bibfnamefont {X.}~\bibnamefont {Jiang}},
  \bibinfo {author} {\bibfnamefont {B.~M.}\ \bibnamefont {Onat}}, \ and\
  \bibinfo {author} {\bibfnamefont {K.}~\bibnamefont {Slomkowski}},\ }in\
  \href@noop {} {\emph {\bibinfo {booktitle} {Quantum Sensing and Nanophotonic
  Devices VII}}},\ Vol.\ \bibinfo {volume} {7608}\ (\bibinfo  {publisher}
  {SPIE},\ \bibinfo {year} {2010})\ p.\ \bibinfo {pages} {760829 (13
  pp.)}\BibitemShut {NoStop}%
\bibitem [{\citenamefont {Knill}, \citenamefont {Laflamme},\ and\ \citenamefont
  {Milburn}(2001)}]{KLM_Nature_01}%
  \BibitemOpen
  \bibfield  {author} {\bibinfo {author} {\bibfnamefont {E.}~\bibnamefont
  {Knill}}, \bibinfo {author} {\bibfnamefont {R.}~\bibnamefont {Laflamme}}, \
  and\ \bibinfo {author} {\bibfnamefont {G.~J.}\ \bibnamefont {Milburn}},\
  }\href@noop {} {\bibfield  {journal} {\bibinfo  {journal} {Nature}\ }\textbf
  {\bibinfo {volume} {409}},\ \bibinfo {pages} {46} (\bibinfo {year}
  {2001})}\BibitemShut {NoStop}%
\bibitem [{\citenamefont {Bennett}\ and\ \citenamefont
  {Brassard}(1984)}]{BB84}%
  \BibitemOpen
  \bibfield  {author} {\bibinfo {author} {\bibfnamefont {C.~H.}\ \bibnamefont
  {Bennett}}\ and\ \bibinfo {author} {\bibfnamefont {G.}~\bibnamefont
  {Brassard}}\ }(\bibinfo {year} {1984})\ pp.\ \bibinfo {pages}
  {175--179}\BibitemShut {NoStop}%
\bibitem [{\citenamefont {Gisin}\ \emph {et~al.}(2002)\citenamefont {Gisin},
  \citenamefont {Ribordy}, \citenamefont {Tittel},\ and\ \citenamefont
  {Zbinden}}]{Gisin_RevModPhy_02_QKD}%
  \BibitemOpen
  \bibfield  {author} {\bibinfo {author} {\bibfnamefont {N.}~\bibnamefont
  {Gisin}}, \bibinfo {author} {\bibfnamefont {G.}~\bibnamefont {Ribordy}},
  \bibinfo {author} {\bibfnamefont {W.}~\bibnamefont {Tittel}}, \ and\ \bibinfo
  {author} {\bibfnamefont {H.}~\bibnamefont {Zbinden}},\ }\href@noop {}
  {\bibfield  {journal} {\bibinfo  {journal} {Rev.Mod.Phys.}\ }\textbf
  {\bibinfo {volume} {74}},\ \bibinfo {pages} {145} (\bibinfo {year}
  {2002})}\BibitemShut {NoStop}%
\bibitem [{\citenamefont {Stellari}\ \emph {et~al.}(2001)\citenamefont
  {Stellari}, \citenamefont {Zappa}, \citenamefont {Cova}, \citenamefont
  {Porta},\ and\ \citenamefont {Tsang}}]{Stellari_TED_01}%
  \BibitemOpen
  \bibfield  {author} {\bibinfo {author} {\bibfnamefont {F.}~\bibnamefont
  {Stellari}}, \bibinfo {author} {\bibfnamefont {F.}~\bibnamefont {Zappa}},
  \bibinfo {author} {\bibfnamefont {S.}~\bibnamefont {Cova}}, \bibinfo {author}
  {\bibfnamefont {C.}~\bibnamefont {Porta}}, \ and\ \bibinfo {author}
  {\bibfnamefont {J.~C.}\ \bibnamefont {Tsang}},\ }\href@noop {} {\bibfield
  {journal} {\bibinfo  {journal} {IEEE Transactions on Electron Devices}\
  }\textbf {\bibinfo {volume} {48}},\ \bibinfo {pages} {2830} (\bibinfo {year}
  {2001})}\BibitemShut {NoStop}%
\bibitem [{\citenamefont {Williams}\ and\ \citenamefont
  {Huntington}(2006)}]{Williams_SPIE_06_linearMode}%
  \BibitemOpen
  \bibfield  {author} {\bibinfo {author} {\bibfnamefont {G.~M.}\ \bibnamefont
  {Williams}}\ and\ \bibinfo {author} {\bibfnamefont {A.~S.}\ \bibnamefont
  {Huntington}},\ }in\ \href@noop {} {\emph {\bibinfo {booktitle} {Spaceborne
  Sensors III}}},\ Vol.\ \bibinfo {volume} {6220}\ (\bibinfo  {publisher}
  {SPIE},\ \bibinfo {year} {2006})\ Chap.~\bibinfo {chapter} {1}, pp.\ \bibinfo
  {pages} {622008--1}\BibitemShut {NoStop}%
\bibitem [{\citenamefont {Tosi}\ \emph {et~al.}(2006)\citenamefont {Tosi},
  \citenamefont {Gallivanoni}, \citenamefont {Zappa},\ and\ \citenamefont
  {Cova}}]{Tosi_SPIE_06}%
  \BibitemOpen
  \bibfield  {author} {\bibinfo {author} {\bibfnamefont {A.}~\bibnamefont
  {Tosi}}, \bibinfo {author} {\bibfnamefont {A.}~\bibnamefont {Gallivanoni}},
  \bibinfo {author} {\bibfnamefont {F.}~\bibnamefont {Zappa}}, \ and\ \bibinfo
  {author} {\bibfnamefont {S.}~\bibnamefont {Cova}},\ }in\ \href@noop {} {\emph
  {\bibinfo {booktitle} {Advanced Photon Counting Techniques}}},\ Vol.\
  \bibinfo {volume} {6372}\ (\bibinfo  {publisher} {SPIE},\ \bibinfo {year}
  {2006})\ pp.\ \bibinfo {pages} {63720--1}\BibitemShut {NoStop}%
\bibitem [{\citenamefont {Thew}\ \emph {et~al.}(2007)\citenamefont {Thew},
  \citenamefont {Stucki}, \citenamefont {Gautier}, \citenamefont {Zbinden},\
  and\ \citenamefont {Rochas}}]{Thew_APL_07}%
  \BibitemOpen
  \bibfield  {author} {\bibinfo {author} {\bibfnamefont {R.~T.}\ \bibnamefont
  {Thew}}, \bibinfo {author} {\bibfnamefont {D.}~\bibnamefont {Stucki}},
  \bibinfo {author} {\bibfnamefont {J.-D.}\ \bibnamefont {Gautier}}, \bibinfo
  {author} {\bibfnamefont {H.}~\bibnamefont {Zbinden}}, \ and\ \bibinfo
  {author} {\bibfnamefont {A.}~\bibnamefont {Rochas}},\ }\href@noop {}
  {\bibfield  {journal} {\bibinfo  {journal} {Applied Physics Letters}\
  }\textbf {\bibinfo {volume} {91}},\ \bibinfo {eid} {201114} (\bibinfo {year}
  {2007})}\BibitemShut {NoStop}%
\bibitem [{\citenamefont {Warburton}, \citenamefont {Itzler},\ and\
  \citenamefont {Buller}(0216)}]{Warburton_09_APL}%
  \BibitemOpen
  \bibfield  {author} {\bibinfo {author} {\bibfnamefont {R.}~\bibnamefont
  {Warburton}}, \bibinfo {author} {\bibfnamefont {M.}~\bibnamefont {Itzler}}, \
  and\ \bibinfo {author} {\bibfnamefont {G.}~\bibnamefont {Buller}},\
  }\href@noop {} {\bibfield  {journal} {\bibinfo  {journal} {Applied Physics
  Letters}\ }\textbf {\bibinfo {volume} {94}},\ \bibinfo {pages} {071116 (3
  pp.)} (\bibinfo {year} {2009/02/16})}\BibitemShut {NoStop}%
\bibitem [{\citenamefont {Zhang}\ \emph {et~al.}(2010)\citenamefont {Zhang},
  \citenamefont {Eraerds}, \citenamefont {Walenta}, \citenamefont {Barreiro},
  \citenamefont {Thew},\ and\ \citenamefont {Zbinden}}]{Zhang_SPIE_10}%
  \BibitemOpen
  \bibfield  {author} {\bibinfo {author} {\bibfnamefont {J.}~\bibnamefont
  {Zhang}}, \bibinfo {author} {\bibfnamefont {P.}~\bibnamefont {Eraerds}},
  \bibinfo {author} {\bibfnamefont {N.}~\bibnamefont {Walenta}}, \bibinfo
  {author} {\bibfnamefont {C.}~\bibnamefont {Barreiro}}, \bibinfo {author}
  {\bibfnamefont {R.}~\bibnamefont {Thew}}, \ and\ \bibinfo {author}
  {\bibfnamefont {H.}~\bibnamefont {Zbinden}}\ }(\bibinfo  {publisher} {SPIE},\
  \bibinfo {year} {2010})\ p.\ \bibinfo {pages} {76810Z}\BibitemShut {NoStop}%
\bibitem [{\citenamefont {Zhao}\ \emph
  {et~al.}(2007{\natexlab{a}})\citenamefont {Zhao}, \citenamefont {Zhang},
  \citenamefont {Lo},\ and\ \citenamefont {Farr}}]{Zhao_APL_07}%
  \BibitemOpen
  \bibfield  {author} {\bibinfo {author} {\bibfnamefont {K.}~\bibnamefont
  {Zhao}}, \bibinfo {author} {\bibfnamefont {A.}~\bibnamefont {Zhang}},
  \bibinfo {author} {\bibfnamefont {Y.-H.}\ \bibnamefont {Lo}}, \ and\ \bibinfo
  {author} {\bibfnamefont {W.}~\bibnamefont {Farr}},\ }\href@noop {} {\bibfield
   {journal} {\bibinfo  {journal} {Appl. Phys. Lett.}\ }\textbf {\bibinfo
  {volume} {91}} (\bibinfo {year} {2007}{\natexlab{a}})}\BibitemShut {NoStop}%
\bibitem [{\citenamefont {Cheng}\ \emph {et~al.}(2011)\citenamefont {Cheng},
  \citenamefont {You}, \citenamefont {Rahman},\ and\ \citenamefont
  {Lo}}]{Cheng_OptExpress_11}%
  \BibitemOpen
  \bibfield  {author} {\bibinfo {author} {\bibfnamefont {J.}~\bibnamefont
  {Cheng}}, \bibinfo {author} {\bibfnamefont {S.}~\bibnamefont {You}}, \bibinfo
  {author} {\bibfnamefont {S.}~\bibnamefont {Rahman}}, \ and\ \bibinfo {author}
  {\bibfnamefont {Y.-H.}\ \bibnamefont {Lo}},\ }\href@noop {} {\bibfield
  {journal} {\bibinfo  {journal} {Optics Express}\ }\textbf {\bibinfo {volume}
  {19}},\ \bibinfo {pages} {15149} (\bibinfo {year} {2011})}\BibitemShut
  {NoStop}%
\bibitem [{\citenamefont {Horsfield}\ \emph {et~al.}(2010)\citenamefont
  {Horsfield}, \citenamefont {Rubery}, \citenamefont {Mack}, \citenamefont
  {Young}, \citenamefont {Herrmann}, \citenamefont {Caldwell}, \citenamefont
  {Evans}, \citenamefont {Sedilleo}, \citenamefont {Kim}, \citenamefont
  {McEvoy}, \citenamefont {Milnes}, \citenamefont {Howorth}, \citenamefont
  {Davis}, \citenamefont {O'Gara}, \citenamefont {Garza}, \citenamefont
  {Miller}, \citenamefont {Stoeffl},\ and\ \citenamefont
  {Ali}}]{Horsfield_RSI_10}%
  \BibitemOpen
  \bibfield  {author} {\bibinfo {author} {\bibfnamefont {C.~J.}\ \bibnamefont
  {Horsfield}}, \bibinfo {author} {\bibfnamefont {M.~S.}\ \bibnamefont
  {Rubery}}, \bibinfo {author} {\bibfnamefont {J.~M.}\ \bibnamefont {Mack}},
  \bibinfo {author} {\bibfnamefont {C.~S.}\ \bibnamefont {Young}}, \bibinfo
  {author} {\bibfnamefont {H.~W.}\ \bibnamefont {Herrmann}}, \bibinfo {author}
  {\bibfnamefont {S.~E.}\ \bibnamefont {Caldwell}}, \bibinfo {author}
  {\bibfnamefont {S.~C.}\ \bibnamefont {Evans}}, \bibinfo {author}
  {\bibfnamefont {T.~J.}\ \bibnamefont {Sedilleo}}, \bibinfo {author}
  {\bibfnamefont {Y.~H.}\ \bibnamefont {Kim}}, \bibinfo {author} {\bibfnamefont
  {A.}~\bibnamefont {McEvoy}}, \bibinfo {author} {\bibfnamefont {J.~S.}\
  \bibnamefont {Milnes}}, \bibinfo {author} {\bibfnamefont {J.}~\bibnamefont
  {Howorth}}, \bibinfo {author} {\bibfnamefont {B.}~\bibnamefont {Davis}},
  \bibinfo {author} {\bibfnamefont {P.~M.}\ \bibnamefont {O'Gara}}, \bibinfo
  {author} {\bibfnamefont {I.}~\bibnamefont {Garza}}, \bibinfo {author}
  {\bibfnamefont {E.~K.}\ \bibnamefont {Miller}}, \bibinfo {author}
  {\bibfnamefont {W.}~\bibnamefont {Stoeffl}}, \ and\ \bibinfo {author}
  {\bibfnamefont {Z.}~\bibnamefont {Ali}},\ }\href@noop {} {\bibfield
  {journal} {\bibinfo  {journal} {Review of Scientific Instruments}\ }\textbf
  {\bibinfo {volume} {81}},\ \bibinfo {pages} {10D318} (\bibinfo {year}
  {2010})}\BibitemShut {NoStop}%
\bibitem [{\citenamefont {Itzler}\ \emph {et~al.}(2007)\citenamefont {Itzler},
  \citenamefont {Jiang}, \citenamefont {Ben-Michael}, \citenamefont
  {Slomkowski}, \citenamefont {Krainak}, \citenamefont {Wu},\ and\
  \citenamefont {Sun}}]{Itzler_SPIE_07}%
  \BibitemOpen
  \bibfield  {author} {\bibinfo {author} {\bibfnamefont {M.~A.}\ \bibnamefont
  {Itzler}}, \bibinfo {author} {\bibfnamefont {X.}~\bibnamefont {Jiang}},
  \bibinfo {author} {\bibfnamefont {R.}~\bibnamefont {Ben-Michael}}, \bibinfo
  {author} {\bibfnamefont {K.}~\bibnamefont {Slomkowski}}, \bibinfo {author}
  {\bibfnamefont {M.~A.}\ \bibnamefont {Krainak}}, \bibinfo {author}
  {\bibfnamefont {S.}~\bibnamefont {Wu}}, \ and\ \bibinfo {author}
  {\bibfnamefont {X.}~\bibnamefont {Sun}},\ }in\ \href@noop {} {\emph {\bibinfo
  {booktitle} {Enabling Photonics Technologies for Defense, Security, and
  Aerospace Applications III}}},\ Vol.\ \bibinfo {volume} {6572}\ (\bibinfo
  {publisher} {SPIE},\ \bibinfo {year} {2007})\ pp.\ \bibinfo {pages}
  {65720--1}\BibitemShut {NoStop}%
\bibitem [{\citenamefont {Albota}\ and\ \citenamefont
  {Wong}(2004)}]{Albota_04_upconversion}%
  \BibitemOpen
  \bibfield  {author} {\bibinfo {author} {\bibfnamefont {M.~A.}\ \bibnamefont
  {Albota}}\ and\ \bibinfo {author} {\bibfnamefont {F.~N.~C.}\ \bibnamefont
  {Wong}},\ }\href@noop {} {\bibfield  {journal} {\bibinfo  {journal} {Optics
  Letters}\ }\textbf {\bibinfo {volume} {29}},\ \bibinfo {pages} {1449}
  (\bibinfo {year} {2004})}\BibitemShut {NoStop}%
\bibitem [{\citenamefont {Lita}, \citenamefont {Miller},\ and\ \citenamefont
  {Nam}(2008)}]{TES:Lita:08}%
  \BibitemOpen
  \bibfield  {author} {\bibinfo {author} {\bibfnamefont {A.~E.}\ \bibnamefont
  {Lita}}, \bibinfo {author} {\bibfnamefont {A.~J.}\ \bibnamefont {Miller}}, \
  and\ \bibinfo {author} {\bibfnamefont {S.~W.}\ \bibnamefont {Nam}},\
  }\href@noop {} {\bibfield  {journal} {\bibinfo  {journal} {Opt. Express}\
  }\textbf {\bibinfo {volume} {16}},\ \bibinfo {pages} {3032} (\bibinfo {year}
  {2008})}\BibitemShut {NoStop}%
\bibitem [{\citenamefont {Gol'tsman}\ \emph {et~al.}(2001)\citenamefont
  {Gol'tsman}, \citenamefont {Okunev}, \citenamefont {Chulkova}, \citenamefont
  {Lipatov}, \citenamefont {Semenov}, \citenamefont {Smirnov}, \citenamefont
  {Voronov}, \citenamefont {Dzardanov}, \citenamefont {Williams},\ and\
  \citenamefont {Sobolewski}}]{Goltsman_APL_01}%
  \BibitemOpen
  \bibfield  {author} {\bibinfo {author} {\bibfnamefont {G.}~\bibnamefont
  {Gol'tsman}}, \bibinfo {author} {\bibfnamefont {O.}~\bibnamefont {Okunev}},
  \bibinfo {author} {\bibfnamefont {G.}~\bibnamefont {Chulkova}}, \bibinfo
  {author} {\bibfnamefont {A.}~\bibnamefont {Lipatov}}, \bibinfo {author}
  {\bibfnamefont {A.}~\bibnamefont {Semenov}}, \bibinfo {author} {\bibfnamefont
  {K.}~\bibnamefont {Smirnov}}, \bibinfo {author} {\bibfnamefont
  {B.}~\bibnamefont {Voronov}}, \bibinfo {author} {\bibfnamefont
  {A.}~\bibnamefont {Dzardanov}}, \bibinfo {author} {\bibfnamefont
  {C.}~\bibnamefont {Williams}}, \ and\ \bibinfo {author} {\bibfnamefont
  {R.}~\bibnamefont {Sobolewski}},\ }\href@noop {} {\bibfield  {journal}
  {\bibinfo  {journal} {Applied Physics Letters}\ }\textbf {\bibinfo {volume}
  {79}},\ \bibinfo {pages} {705} (\bibinfo {year} {2001})}\BibitemShut
  {NoStop}%
\bibitem [{\citenamefont {Yan}, \citenamefont {Majedi},\ and\ \citenamefont
  {Safavi-Naeini}(2007)}]{zyan_TAS_07}%
  \BibitemOpen
  \bibfield  {author} {\bibinfo {author} {\bibfnamefont {Z.}~\bibnamefont
  {Yan}}, \bibinfo {author} {\bibfnamefont {A.}~\bibnamefont {Majedi}}, \ and\
  \bibinfo {author} {\bibfnamefont {S.}~\bibnamefont {Safavi-Naeini}},\
  }\href@noop {} {\bibfield  {journal} {\bibinfo  {journal} {Applied
  Superconductivity, IEEE Transactions on}\ }\textbf {\bibinfo {volume} {17}},\
  \bibinfo {pages} {3789 } (\bibinfo {year} {2007})}\BibitemShut {NoStop}%
\bibitem [{\citenamefont {Anant}\ \emph {et~al.}(2008)\citenamefont {Anant},
  \citenamefont {Kerman}, \citenamefont {Dauler}, \citenamefont {Yang},
  \citenamefont {Rosfjord},\ and\ \citenamefont
  {Berggren}}]{Anant_OptExpress_08}%
  \BibitemOpen
  \bibfield  {author} {\bibinfo {author} {\bibfnamefont {V.}~\bibnamefont
  {Anant}}, \bibinfo {author} {\bibfnamefont {A.~J.}\ \bibnamefont {Kerman}},
  \bibinfo {author} {\bibfnamefont {E.~A.}\ \bibnamefont {Dauler}}, \bibinfo
  {author} {\bibfnamefont {J.~K.~W.}\ \bibnamefont {Yang}}, \bibinfo {author}
  {\bibfnamefont {K.~M.}\ \bibnamefont {Rosfjord}}, \ and\ \bibinfo {author}
  {\bibfnamefont {K.~K.}\ \bibnamefont {Berggren}},\ }\href@noop {} {\bibfield
  {journal} {\bibinfo  {journal} {Opt. Express}\ }\textbf {\bibinfo {volume}
  {16}},\ \bibinfo {pages} {10750} (\bibinfo {year} {2008})}\BibitemShut
  {NoStop}%
\bibitem [{\citenamefont {Zhao}\ \emph
  {et~al.}(2007{\natexlab{b}})\citenamefont {Zhao}, \citenamefont {Zhang},
  \citenamefont {hwa Lo},\ and\ \citenamefont {Farr}}]{Zhao_APL_0701}%
  \BibitemOpen
  \bibfield  {author} {\bibinfo {author} {\bibfnamefont {K.}~\bibnamefont
  {Zhao}}, \bibinfo {author} {\bibfnamefont {A.}~\bibnamefont {Zhang}},
  \bibinfo {author} {\bibfnamefont {Y.}~\bibnamefont {hwa Lo}}, \ and\ \bibinfo
  {author} {\bibfnamefont {W.}~\bibnamefont {Farr}},\ }\href@noop {} {\bibfield
   {journal} {\bibinfo  {journal} {Applied Physics Letters}\ }\textbf {\bibinfo
  {volume} {91}},\ \bibinfo {eid} {081107} (\bibinfo {year}
  {2007}{\natexlab{b}})}\BibitemShut {NoStop}%
\bibitem [{\citenamefont {Zhao}\ \emph {et~al.}(2008)\citenamefont {Zhao},
  \citenamefont {You}, \citenamefont {Cheng},\ and\ \citenamefont {hwa
  Lo}}]{Zhao_APL_08}%
  \BibitemOpen
  \bibfield  {author} {\bibinfo {author} {\bibfnamefont {K.}~\bibnamefont
  {Zhao}}, \bibinfo {author} {\bibfnamefont {S.}~\bibnamefont {You}}, \bibinfo
  {author} {\bibfnamefont {J.}~\bibnamefont {Cheng}}, \ and\ \bibinfo {author}
  {\bibfnamefont {Y.}~\bibnamefont {hwa Lo}},\ }\href@noop {} {\bibfield
  {journal} {\bibinfo  {journal} {Applied Physics Letters}\ }\textbf {\bibinfo
  {volume} {93}},\ \bibinfo {eid} {153504} (\bibinfo {year}
  {2008})}\BibitemShut {NoStop}%
\bibitem [{\citenamefont {Hayat}\ \emph {et~al.}(2010)\citenamefont {Hayat},
  \citenamefont {Itzler}, \citenamefont {Ramirez},\ and\ \citenamefont
  {Rees}}]{Hayat_NFAD_model_10}%
  \BibitemOpen
  \bibfield  {author} {\bibinfo {author} {\bibfnamefont {M.~M.}\ \bibnamefont
  {Hayat}}, \bibinfo {author} {\bibfnamefont {M.~A.}\ \bibnamefont {Itzler}},
  \bibinfo {author} {\bibfnamefont {D.~A.}\ \bibnamefont {Ramirez}}, \ and\
  \bibinfo {author} {\bibfnamefont {G.~J.}\ \bibnamefont {Rees}},\ }in\
  \href@noop {} {\emph {\bibinfo {booktitle} {Quantum Sensing and Nanophotonic
  Devices VII}}},\ Vol.\ \bibinfo {volume} {7608}\ (\bibinfo  {publisher}
  {SPIE},\ \bibinfo {year} {2010})\ p.\ \bibinfo {pages} {76082B (8
  pp.)}\BibitemShut {NoStop}%
\bibitem [{\citenamefont {Jiang}\ \emph
  {et~al.}(2009{\natexlab{b}})\citenamefont {Jiang}, \citenamefont {Itzler},
  \citenamefont {Nyman},\ and\ \citenamefont {Slomkowski}}]{Itzler_SPIE09}%
  \BibitemOpen
  \bibfield  {author} {\bibinfo {author} {\bibfnamefont {X.}~\bibnamefont
  {Jiang}}, \bibinfo {author} {\bibfnamefont {M.~A.}\ \bibnamefont {Itzler}},
  \bibinfo {author} {\bibfnamefont {B.}~\bibnamefont {Nyman}}, \ and\ \bibinfo
  {author} {\bibfnamefont {K.}~\bibnamefont {Slomkowski}},\ }in\ \href@noop {}
  {\emph {\bibinfo {booktitle} {Advanced Photon Counting Techniques III}}},\
  Vol.\ \bibinfo {volume} {7320}\ (\bibinfo  {publisher} {SPIE},\ \bibinfo
  {year} {2009})\ p.\ \bibinfo {pages} {732011 (10 pp.)}\BibitemShut {NoStop}%
\bibitem [{\citenamefont {Jennewein}, \citenamefont {Barbieri},\ and\
  \citenamefont {White}(2011)}]{Jennewein_JMO_11}%
  \BibitemOpen
  \bibfield  {author} {\bibinfo {author} {\bibfnamefont {T.}~\bibnamefont
  {Jennewein}}, \bibinfo {author} {\bibfnamefont {M.}~\bibnamefont {Barbieri}},
  \ and\ \bibinfo {author} {\bibfnamefont {A.~G.}\ \bibnamefont {White}},\
  }\href@noop {} {\bibfield  {journal} {\bibinfo  {journal} {Journal of Modern
  Optics}\ }\textbf {\bibinfo {volume} {58}},\ \bibinfo {pages} {276} (\bibinfo
  {year} {2011})}\BibitemShut {NoStop}%
\bibitem [{\citenamefont {Lundeen}\ \emph {et~al.}(2009)\citenamefont
  {Lundeen}, \citenamefont {Feito}, \citenamefont {Coldenstrodt-Ronge},
  \citenamefont {Pregnell}, \citenamefont {Ralph}, \citenamefont {Silberhorn},
  \citenamefont {Eisert}, \citenamefont {Plenio},\ and\ \citenamefont
  {Walmsley}}]{Lundeen_NatPhy_09}%
  \BibitemOpen
  \bibfield  {author} {\bibinfo {author} {\bibfnamefont {J.}~\bibnamefont
  {Lundeen}}, \bibinfo {author} {\bibfnamefont {A.}~\bibnamefont {Feito}},
  \bibinfo {author} {\bibfnamefont {H.}~\bibnamefont {Coldenstrodt-Ronge}},
  \bibinfo {author} {\bibfnamefont {K.}~\bibnamefont {Pregnell}}, \bibinfo
  {author} {\bibfnamefont {C.}~\bibnamefont {Ralph}}, \bibinfo {author}
  {\bibfnamefont {C.}~\bibnamefont {Silberhorn}}, \bibinfo {author}
  {\bibfnamefont {J.}~\bibnamefont {Eisert}}, \bibinfo {author} {\bibfnamefont
  {M.}~\bibnamefont {Plenio}}, \ and\ \bibinfo {author} {\bibfnamefont
  {I.}~\bibnamefont {Walmsley}},\ }\href@noop {} {\bibfield  {journal}
  {\bibinfo  {journal} {Nature Physics}\ }\textbf {\bibinfo {volume} {5}},\
  \bibinfo {pages} {27} (\bibinfo {year} {2009})}\BibitemShut {NoStop}%
\bibitem [{\citenamefont {Akhlaghi}, \citenamefont {Majedi},\ and\
  \citenamefont {Lundeen}(2011)}]{Akhlaghi_OptExp_11}%
  \BibitemOpen
  \bibfield  {author} {\bibinfo {author} {\bibfnamefont {M.}~\bibnamefont
  {Akhlaghi}}, \bibinfo {author} {\bibfnamefont {A.}~\bibnamefont {Majedi}}, \
  and\ \bibinfo {author} {\bibfnamefont {J.}~\bibnamefont {Lundeen}},\
  }\href@noop {} {\bibfield  {journal} {\bibinfo  {journal} {Optics Express}\
  }\textbf {\bibinfo {volume} {19}},\ \bibinfo {pages} {21305} (\bibinfo {year}
  {2011})}\BibitemShut {NoStop}%
\bibitem [{Rep({\natexlab{a}})}]{Reply_1a}%
  \BibitemOpen
  \href@noop {} {}\bibinfo {note} {To determine the attenuation settings, we
  use a commercial power meter (OZ-300) to do the calibration. We first set a
  proper repletion rate $f_\mathrm{rep}$ to trigger the laser, and then set a
  fixed attenuation value $t_a^0$ to the variable attenuator (VOA) to make sure
  the VOA output power $P_0$ is well above the noise limit of the power meter.
  Then the relation between average photon number $n(x)$ to the attenuator
  setting $x$ is $ n\left( x \right) = {{10^{{{P_0 + t_a^0 - x} \over {10}}}
  \lambda } \over {1000h c f_\mathrm{rep} }} $, where $\lambda$ is the
  wavelength, $h$ is the Plank constant, $c$ is the free space speed of light.
  Then we can solve for any arbitrary photon number $n$ with respect to
  attenuator setting value $x$}\BibitemShut {NoStop}%
\bibitem [{Rep({\natexlab{b}})}]{Reply_1b}%
  \BibitemOpen
  \href@noop {} {}\bibinfo {note} {Three potential problems of the commercial
  id-201 detector module must be taken into account: (1) its detector
  temperature is monitored by the firmware; (2) we carefully adjust the delay
  of the arrival of the photons to make sure photon detection is within the
  right value; (3) the fiber connector has been carefully cleaned and inspected
  by the fiber scope for all of our measurements}\BibitemShut {NoStop}%
\bibitem [{\citenamefont {Mollow}\ and\ \citenamefont
  {Glauber}(1967)}]{Mollow_PhysRev_67_I}%
  \BibitemOpen
  \bibfield  {author} {\bibinfo {author} {\bibfnamefont {B.~R.}\ \bibnamefont
  {Mollow}}\ and\ \bibinfo {author} {\bibfnamefont {R.~J.}\ \bibnamefont
  {Glauber}},\ }\href@noop {} {\bibfield  {journal} {\bibinfo  {journal} {Phys.
  Rev.}\ }\textbf {\bibinfo {volume} {160}},\ \bibinfo {pages} {1076} (\bibinfo
  {year} {1967})}\BibitemShut {NoStop}%
\bibitem [{\citenamefont {Ware}\ and\ \citenamefont
  {Migdall}(2004)}]{Migdall_JMO_04}%
  \BibitemOpen
  \bibfield  {author} {\bibinfo {author} {\bibfnamefont {M.}~\bibnamefont
  {Ware}}\ and\ \bibinfo {author} {\bibfnamefont {A.}~\bibnamefont {Migdall}},\
  }\href@noop {} {\bibfield  {journal} {\bibinfo  {journal} {Journal of Modern
  Optics}\ }\textbf {\bibinfo {volume} {51}},\ \bibinfo {pages} {1549}
  (\bibinfo {year} {2004})}\BibitemShut {NoStop}%
\bibitem [{\citenamefont {Burnham}\ and\ \citenamefont
  {Weinberg}(1970)}]{Burnham:70}%
  \BibitemOpen
  \bibfield  {author} {\bibinfo {author} {\bibfnamefont {D.~C.}\ \bibnamefont
  {Burnham}}\ and\ \bibinfo {author} {\bibfnamefont {D.~L.}\ \bibnamefont
  {Weinberg}},\ }\href@noop {} {\bibfield  {journal} {\bibinfo  {journal}
  {Phys. Rev. Lett.}\ }\textbf {\bibinfo {volume} {25}},\ \bibinfo {pages} {84}
  (\bibinfo {year} {1970})}\BibitemShut {NoStop}%
\bibitem [{\citenamefont {Polyakov}\ and\ \citenamefont
  {Migdall}(2007)}]{Polyakov:07}%
  \BibitemOpen
  \bibfield  {author} {\bibinfo {author} {\bibfnamefont {S.~V.}\ \bibnamefont
  {Polyakov}}\ and\ \bibinfo {author} {\bibfnamefont {A.~L.}\ \bibnamefont
  {Migdall}},\ }\href@noop {} {\bibfield  {journal} {\bibinfo  {journal} {Opt.
  Express}\ }\textbf {\bibinfo {volume} {15}},\ \bibinfo {pages} {1390}
  (\bibinfo {year} {2007})}\BibitemShut {NoStop}%
\bibitem [{\citenamefont {Kwiat}\ \emph {et~al.}(1994)\citenamefont {Kwiat},
  \citenamefont {Steinberg}, \citenamefont {Chiao}, \citenamefont {Eberhard},\
  and\ \citenamefont {Petroff}}]{Kwiat:94}%
  \BibitemOpen
  \bibfield  {author} {\bibinfo {author} {\bibfnamefont {P.~G.}\ \bibnamefont
  {Kwiat}}, \bibinfo {author} {\bibfnamefont {A.~M.}\ \bibnamefont
  {Steinberg}}, \bibinfo {author} {\bibfnamefont {R.~Y.}\ \bibnamefont
  {Chiao}}, \bibinfo {author} {\bibfnamefont {P.~H.}\ \bibnamefont {Eberhard}},
  \ and\ \bibinfo {author} {\bibfnamefont {M.~D.}\ \bibnamefont {Petroff}},\
  }\href@noop {} {\bibfield  {journal} {\bibinfo  {journal} {Appl. Opt.}\
  }\textbf {\bibinfo {volume} {33}},\ \bibinfo {pages} {1844} (\bibinfo {year}
  {1994})}\BibitemShut {NoStop}%
\bibitem [{\citenamefont {Hadfield}\ \emph {et~al.}(2007)\citenamefont
  {Hadfield}, \citenamefont {Stevens}, \citenamefont {Mirin},\ and\
  \citenamefont {Nam}}]{Hadfield_JAP_07}%
  \BibitemOpen
  \bibfield  {author} {\bibinfo {author} {\bibfnamefont {R.~H.}\ \bibnamefont
  {Hadfield}}, \bibinfo {author} {\bibfnamefont {M.~J.}\ \bibnamefont
  {Stevens}}, \bibinfo {author} {\bibfnamefont {R.~P.}\ \bibnamefont {Mirin}},
  \ and\ \bibinfo {author} {\bibfnamefont {S.~W.}\ \bibnamefont {Nam}},\
  }\href@noop {} {\bibfield  {journal} {\bibinfo  {journal} {J. Appl. Phys.}\
  }\textbf {\bibinfo {volume} {101}},\ \bibinfo {pages} {103104} (\bibinfo
  {year} {2007})}\BibitemShut {NoStop}%
\bibitem [{\citenamefont {Tanzilli}\ \emph {et~al.}(2002)\citenamefont
  {Tanzilli}, \citenamefont {Tittel}, \citenamefont {De~Riedmatten},
  \citenamefont {Zbinden}, \citenamefont {Baldi}, \citenamefont {DeMicheli},
  \citenamefont {Ostrowsky},\ and\ \citenamefont {Gisin}}]{Tanzilli_PPLN_02}%
  \BibitemOpen
  \bibfield  {author} {\bibinfo {author} {\bibfnamefont {S.}~\bibnamefont
  {Tanzilli}}, \bibinfo {author} {\bibfnamefont {W.}~\bibnamefont {Tittel}},
  \bibinfo {author} {\bibfnamefont {H.}~\bibnamefont {De~Riedmatten}}, \bibinfo
  {author} {\bibfnamefont {H.}~\bibnamefont {Zbinden}}, \bibinfo {author}
  {\bibfnamefont {P.}~\bibnamefont {Baldi}}, \bibinfo {author} {\bibfnamefont
  {M.}~\bibnamefont {DeMicheli}}, \bibinfo {author} {\bibfnamefont
  {D.}~\bibnamefont {Ostrowsky}}, \ and\ \bibinfo {author} {\bibfnamefont
  {N.}~\bibnamefont {Gisin}},\ }\href@noop {} {\bibfield  {journal} {\bibinfo
  {journal} {The European Physical Journal D - Atomic, Molecular, Optical and
  Plasma Physics}\ }\textbf {\bibinfo {volume} {18}},\ \bibinfo {pages} {155}
  (\bibinfo {year} {2002})}\BibitemShut {NoStop}%
\bibitem [{\citenamefont {Hadfield}(2009)}]{Hadfield_NatPhoto_09}%
  \BibitemOpen
  \bibfield  {author} {\bibinfo {author} {\bibfnamefont {R.~H.}\ \bibnamefont
  {Hadfield}},\ }\href@noop {} {\bibfield  {journal} {\bibinfo  {journal} {Nat
  Photon}\ }\textbf {\bibinfo {volume} {3}},\ \bibinfo {pages} {696} (\bibinfo
  {year} {2009})}\BibitemShut {NoStop}%
\bibitem [{\citenamefont {Tosi}\ \emph {et~al.}(2009)\citenamefont {Tosi},
  \citenamefont {Mora}, \citenamefont {Zappa},\ and\ \citenamefont
  {Cova}}]{Tosi_JMO_09}%
  \BibitemOpen
  \bibfield  {author} {\bibinfo {author} {\bibfnamefont {A.}~\bibnamefont
  {Tosi}}, \bibinfo {author} {\bibfnamefont {A.~D.}\ \bibnamefont {Mora}},
  \bibinfo {author} {\bibfnamefont {F.}~\bibnamefont {Zappa}}, \ and\ \bibinfo
  {author} {\bibfnamefont {S.}~\bibnamefont {Cova}},\ }\href@noop {} {\bibfield
   {journal} {\bibinfo  {journal} {Journal of Modern Optics}\ }\textbf
  {\bibinfo {volume} {56}},\ \bibinfo {pages} {299} (\bibinfo {year}
  {2009})}\BibitemShut {NoStop}%
\bibitem [{Rep({\natexlab{c}})}]{Reply_6}%
  \BibitemOpen
  \href@noop {} {}\bibinfo {note} {The collection of data using constant
  darkcount rate about 100~CPS because it will guarantee a constant bottom line
  for all different measurement conditions, such as different
  temperatures.}\BibitemShut {Stop}%
\bibitem [{\citenamefont {Jiang}\ \emph {et~al.}(2008)\citenamefont {Jiang},
  \citenamefont {Itzler}, \citenamefont {Ben-Michael}, \citenamefont
  {Slomkowski}, \citenamefont {Krainak}, \citenamefont {Wu},\ and\
  \citenamefont {Sun}}]{Jiang_08_JQE_afterpulsing}%
  \BibitemOpen
  \bibfield  {author} {\bibinfo {author} {\bibfnamefont {X.}~\bibnamefont
  {Jiang}}, \bibinfo {author} {\bibfnamefont {M.~A.}\ \bibnamefont {Itzler}},
  \bibinfo {author} {\bibfnamefont {R.}~\bibnamefont {Ben-Michael}}, \bibinfo
  {author} {\bibfnamefont {K.}~\bibnamefont {Slomkowski}}, \bibinfo {author}
  {\bibfnamefont {M.~A.}\ \bibnamefont {Krainak}}, \bibinfo {author}
  {\bibfnamefont {S.}~\bibnamefont {Wu}}, \ and\ \bibinfo {author}
  {\bibfnamefont {X.}~\bibnamefont {Sun}},\ }\href@noop {} {\bibfield
  {journal} {\bibinfo  {journal} {IEEE Journal of Quantum Electronics}\
  }\textbf {\bibinfo {volume} {44}},\ \bibinfo {pages} {3} (\bibinfo {year}
  {2008})}\BibitemShut {NoStop}%
\bibitem [{Rep({\natexlab{d}})}]{Reply_8a}%
  \BibitemOpen
  \href@noop {} {}\bibinfo {note} {Our setup measures the FWHM over the system
  level instrument response function (IRF) and combines the contributions from
  laser pulse jitter, electronics jitter, and NFAD detector jitter. In fact, we
  have only determined the upper limit of the NFAD timing jitter.}\BibitemShut
  {Stop}%
\bibitem [{\citenamefont {Lo}, \citenamefont {Ma},\ and\ \citenamefont
  {Chen}(2005)}]{Lo:PhysRevLett_Decoy05}%
  \BibitemOpen
  \bibfield  {author} {\bibinfo {author} {\bibfnamefont {H.-K.}\ \bibnamefont
  {Lo}}, \bibinfo {author} {\bibfnamefont {X.}~\bibnamefont {Ma}}, \ and\
  \bibinfo {author} {\bibfnamefont {K.}~\bibnamefont {Chen}},\ }\href@noop {}
  {\bibfield  {journal} {\bibinfo  {journal} {Phys. Rev. Lett.}\ }\textbf
  {\bibinfo {volume} {94}},\ \bibinfo {pages} {230504} (\bibinfo {year}
  {2005})}\BibitemShut {NoStop}%
\bibitem [{\citenamefont {Scherer}\ \emph {et~al.}(2009)\citenamefont
  {Scherer}, \citenamefont {Howard}, \citenamefont {Sanders},\ and\
  \citenamefont {Tittel}}]{Scherer_PRA_09}%
  \BibitemOpen
  \bibfield  {author} {\bibinfo {author} {\bibfnamefont {A.}~\bibnamefont
  {Scherer}}, \bibinfo {author} {\bibfnamefont {R.~B.}\ \bibnamefont {Howard}},
  \bibinfo {author} {\bibfnamefont {B.~C.}\ \bibnamefont {Sanders}}, \ and\
  \bibinfo {author} {\bibfnamefont {W.}~\bibnamefont {Tittel}},\ }\href@noop {}
  {\bibfield  {journal} {\bibinfo  {journal} {Phys. Rev. A}\ }\textbf {\bibinfo
  {volume} {80}},\ \bibinfo {pages} {062310} (\bibinfo {year}
  {2009})}\BibitemShut {NoStop}%
\bibitem [{\citenamefont {Scherer}, \citenamefont {Sanders},\ and\
  \citenamefont {Tittel}(2011)}]{Scherer_OptExpress_11}%
  \BibitemOpen
  \bibfield  {author} {\bibinfo {author} {\bibfnamefont {A.}~\bibnamefont
  {Scherer}}, \bibinfo {author} {\bibfnamefont {B.~C.}\ \bibnamefont
  {Sanders}}, \ and\ \bibinfo {author} {\bibfnamefont {W.}~\bibnamefont
  {Tittel}},\ }\href@noop {} {\bibfield  {journal} {\bibinfo  {journal} {Opt.
  Express}\ }\textbf {\bibinfo {volume} {19}},\ \bibinfo {pages} {3004}
  (\bibinfo {year} {2011})}\BibitemShut {NoStop}%
\bibitem [{\citenamefont {Ma}, \citenamefont {Fung},\ and\ \citenamefont
  {Lo}(2007)}]{Ma_PRA_07}%
  \BibitemOpen
  \bibfield  {author} {\bibinfo {author} {\bibfnamefont {X.}~\bibnamefont
  {Ma}}, \bibinfo {author} {\bibfnamefont {C.-H.~F.}\ \bibnamefont {Fung}}, \
  and\ \bibinfo {author} {\bibfnamefont {H.-K.}\ \bibnamefont {Lo}},\
  }\href@noop {} {\bibfield  {journal} {\bibinfo  {journal} {Phys. Rev. A}\
  }\textbf {\bibinfo {volume} {76}},\ \bibinfo {pages} {012307} (\bibinfo
  {year} {2007})}\BibitemShut {NoStop}%
\bibitem [{\citenamefont {Scheidl}\ \emph {et~al.}(2009)\citenamefont
  {Scheidl}, \citenamefont {Ursin}, \citenamefont {Fedrizzi}, \citenamefont
  {Ramelow}, \citenamefont {Ma}, \citenamefont {Herbst}, \citenamefont
  {Prevedel}, \citenamefont {Ratschbacher}, \citenamefont {Kofler},
  \citenamefont {Jennewein},\ and\ \citenamefont {Zeilinger}}]{Scheidl_NJP_09}%
  \BibitemOpen
  \bibfield  {author} {\bibinfo {author} {\bibfnamefont {T.}~\bibnamefont
  {Scheidl}}, \bibinfo {author} {\bibfnamefont {R.}~\bibnamefont {Ursin}},
  \bibinfo {author} {\bibfnamefont {A.}~\bibnamefont {Fedrizzi}}, \bibinfo
  {author} {\bibfnamefont {S.}~\bibnamefont {Ramelow}}, \bibinfo {author}
  {\bibfnamefont {X.-S.}\ \bibnamefont {Ma}}, \bibinfo {author} {\bibfnamefont
  {T.}~\bibnamefont {Herbst}}, \bibinfo {author} {\bibfnamefont
  {R.}~\bibnamefont {Prevedel}}, \bibinfo {author} {\bibfnamefont
  {L.}~\bibnamefont {Ratschbacher}}, \bibinfo {author} {\bibfnamefont
  {J.}~\bibnamefont {Kofler}}, \bibinfo {author} {\bibfnamefont
  {T.}~\bibnamefont {Jennewein}}, \ and\ \bibinfo {author} {\bibfnamefont
  {A.}~\bibnamefont {Zeilinger}},\ }\href@noop {} {\bibfield  {journal}
  {\bibinfo  {journal} {New Journal of Physics}\ }\textbf {\bibinfo {volume}
  {11}},\ \bibinfo {pages} {085002} (\bibinfo {year} {2009})}\BibitemShut
  {NoStop}%
\bibitem [{Rep({\natexlab{e}})}]{Reply_Eq3}%
  \BibitemOpen
  \href@noop {} {}\bibinfo {note} {To derive our expression for the QBER, we
  assume $Q_{BER} = {1 \over 2}\left( {1 - \mathrm{Vis}} \right)$, and further
  assume that the entanglement visibility is $\mathrm{Vis} = {{\max - \min }
  \over {\max + \min }}$. The coincidence counts are estimated by $N_c = N\eta
  _A \eta _B$; The accidental counts are expressed by $N_a = \left[ {N\eta _A
  \left( {1 + \kappa _A } \right) + D_A } \right]\left[ {N\eta _b \left( {1 +
  \kappa _B } \right) + D_B } \right]$; and consequently we have $\max = N_a +
  N_c$, and $\min = N_a$. We compared the QBER given by Eq. (\ref{eq:QBER})
  with the results in Ref. \onlinecite{Ma_PRA_07} to confirm that they both
  asymptotically agree for large losses assuming that both detectors are ideal
  and do not show any afterpulsing.}\BibitemShut {Stop}%
\bibitem [{dot()}]{dotfast}%
  \BibitemOpen
  \href@noop {} {}\bibinfo {note} {Available at
  http://www.dotfast-consulting.at/; and http://www.uqdevices.com/}\BibitemShut
  {NoStop}%
\bibitem [{\citenamefont {Stucki}\ \emph {et~al.}(2009)\citenamefont {Stucki},
  \citenamefont {Walenta}, \citenamefont {Vannel}, \citenamefont {Thew},
  \citenamefont {Gisin}, \citenamefont {Zbinden}, \citenamefont {Gray},
  \citenamefont {Towery},\ and\ \citenamefont {Ten}}]{Stucki_NJP_09}%
  \BibitemOpen
  \bibfield  {author} {\bibinfo {author} {\bibfnamefont {D.}~\bibnamefont
  {Stucki}}, \bibinfo {author} {\bibfnamefont {N.}~\bibnamefont {Walenta}},
  \bibinfo {author} {\bibfnamefont {F.}~\bibnamefont {Vannel}}, \bibinfo
  {author} {\bibfnamefont {R.~T.}\ \bibnamefont {Thew}}, \bibinfo {author}
  {\bibfnamefont {N.}~\bibnamefont {Gisin}}, \bibinfo {author} {\bibfnamefont
  {H.}~\bibnamefont {Zbinden}}, \bibinfo {author} {\bibfnamefont
  {S.}~\bibnamefont {Gray}}, \bibinfo {author} {\bibfnamefont {C.~R.}\
  \bibnamefont {Towery}}, \ and\ \bibinfo {author} {\bibfnamefont
  {S.}~\bibnamefont {Ten}},\ }\href@noop {} {\bibfield  {journal} {\bibinfo
  {journal} {New Journal of Physics}\ }\textbf {\bibinfo {volume} {11}},\
  \bibinfo {pages} {075003} (\bibinfo {year} {2009})}\BibitemShut {NoStop}%
\end{thebibliography}
%

\end{document}